\documentclass[english,prd,superscriptaddress,nofootinbib,preprintnumbers]{revtex4}
\usepackage[latin1]{inputenc}
\usepackage{graphicx}
\usepackage{amsmath,amsthm,amssymb}
\usepackage{bm}

\makeatletter
\@ifundefined{textcolor}{}
{%
 \definecolor{BLACK}{gray}{0}
 \definecolor{WHITE}{gray}{1}
 \definecolor{RED}{rgb}{1,0,0}
 \definecolor{GREEN}{rgb}{0,1,0}
 \definecolor{BLUE}{rgb}{0,0,1}
 \definecolor{CYAN}{cmyk}{1,0,0,0}
 \definecolor{MAGENTA}{cmyk}{0,1,0,0}
 \definecolor{YELLOW}{cmyk}{0,0,1,0}
}

\def\be{\begin{equation}}
\def\ee{\end{equation}}
\def\ba{\begin{eqnarray}}
\def\ea{\end{eqnarray}}

\usepackage{babel}

\makeatother

\usepackage{babel}
\begin{document}

\title{Screening the fifth force in the Horndeski's most general 
scalar-tensor theories}

\author{Ryotaro Kase}
\affiliation{Department of Physics, Faculty of Science, Tokyo University of Science,
1-3, Kagurazaka, Shinjuku-ku, Tokyo 162-8601, Japan}

\author{Shinji Tsujikawa}
\affiliation{Department of Physics, Faculty of Science, Tokyo University of Science,
1-3, Kagurazaka, Shinjuku-ku, Tokyo 162-8601, Japan}

\begin{abstract}
We study how the Vainshtein mechanism operates in the most general 
scalar-tensor theories with second-order equations of motion. 
The field equations of motion, which can be also applicable to most of 
other screening scenarios proposed in literature, 
are generally derived in a spherically 
symmetric space-time with a matter source. 
In the presence of a field coupling to the Ricci scalar,
we clarify conditions under which the Vainshtein 
mechanism is at work in a weak gravitational background.
We also obtain the solutions of the field equation inside 
a spherically symmetric body and show how they can be 
connected to exterior solutions that accommodate the Vainshtein mechanism. 
We apply our general results to a number of concrete 
models such as the covariant/extended Galileons 
and the DBI Galileons with Gauss-Bonnet and other terms. 
In these models the fifth force can be suppressed 
to be compatible with solar-system constraints, 
provided that non-linear field kinetic terms coupled to 
the Einstein tensor do not dominate 
over other non-linear field self-interactions.
\end{abstract}

\date{\today}

\maketitle

\section{Introduction}

Motivated by the dark energy problem, there have been numerous 
attempts to modify General Relativity (GR) at large 
distances (see Refs.~\cite{review} for reviews).
Generally, such modifications give rise to new degrees of freedom 
associated with the breaking of gauge symmetries of GR. 
In $f(R)$ gravity, for example, the Lagrangian including the 
non-linear terms of the Ricci scalar $R$ brings a scalar degree 
of freedom called ``scalarons'' in the gravitational sector \cite{Star80}. 
Those scalar degrees of freedom can freely propagate to 
mediate a long-range fifth force with baryonic matter.
Since the gravitational experiments within the solar system agree 
with GR in high precision, we need to screen the fifth force 
at small distances while realizing the cosmic 
acceleration on large scales.

In modified gravitational theories there are several different ways to
recover the General Relativistic behavior 
in local regions. 
One is the so-called chameleon 
mechanism \cite{chame}, 
under which the effective mass 
$m_{\rm eff}(\phi)$ of a scalar field $\phi$ is 
different depending on the surrounding matter density. 
In the regions of high density with large 
$m_{\rm eff}(\phi)$, 
a spherically symmetric body can have a thin-shell to suppress
the coupling between the field and non-relativistic matter 
outside the body. In fact, the chameleon mechanism was
applied to dark energy models based on $f(R)$ 
theories \cite{fRlocal} and Brans-Dicke theories \cite{Yoko}. 
There exists a similar screening scenario called the symmetron 
mechanism \cite{symmetron}.
The choices of the field potential and the matter coupling 
for symmetrons are different from those for chameleons.
Unfortunately, the energy scale of the simplest potential 
of symmetrons is too small to act for dark energy.

While the existence of the field potentials is crucial for the success of 
the chameleon and symmetron mechanisms, there is another screening 
scenario called the Vainshtein mechanism \cite{Vainshtein} based on derivative 
self-interactions of a scalar degree of freedom. 
The Vainshtein mechanism was originally discovered in the 
context of massive gravity (spin-2 Pauli-Fierz theory \cite{Pauli}).
The helicity-0 mode of massive gravitons does not decouple from 
matter in a linear approximation \cite{DVZ}, 
but the derivative self-interactions of the helicity-0 mode allow 
to suppress the matter coupling \cite{Vainshtein}.
The Vainshtein mechanism can be at work in the 
Dvail-Gabadadze-Porrati (DGP)
braneworld model \cite{DGP} where the cosmic acceleration is realized by 
a gravitational leakage to the extra dimension. 
In the DGP model, a non-linear field self-interaction of the form 
$(\partial \phi)^2\Box \phi$, which arises due to the mixture of 
the longitudinal and transverse gravitons, can lead to the recovery 
of GR within the so-called Vainshtein radius 
$r_V$ \cite{DGPnon,DDGV,Nico,Silva}. 

In flat (Minkowski) space-time, the non-linear field Lagrangian 
$(\partial \phi)^2\Box \phi$ gives rise to the field equation of motion 
respecting the Galilean symmetry $\partial_{\mu} \phi \to \partial_{\mu} \phi+b_{\mu}$. 
Imposing this symmetry in flat space-time, 
the field Lagrangian 
is restricted to have only five terms including $X \equiv -(\partial \phi)^2/2$ 
and $X \Box \phi$ \cite{Nicolis}.
The covariant generalization of this ``Galileon'' theory in curved 
space-time was carried out in Refs.~\cite{Deffayet}.
The Lagrangian of the covariant Galileon is constructed to keep
the equations of motion at second order, while recovering 
the Galilean symmetry in the limit of Minkowski space-time. 
The application of covariant Galileon theory to dark energy 
has been extensively studied in Refs.~\cite{GALcosmo,DTPRL}.

If we consider a probe brane embedded in a five-dimensional 
Minkowski bulk, all the non-linear self-interactions of Galileons naturally 
arise from the brane tension, 
induced curvature, and 
the Gibbons-Hawking-York boundary terms of the bulk 
contributions \cite{Rham}. 
Moreover, the coupling to gravity is straightforward by taking the 
induced metric on the brane in the form 
$g_{\mu \nu}=q_{\mu \nu}+\partial_{\mu} \phi \partial_{\nu} \phi$, 
where $q_{\mu \nu}$ is an arbitrary four-dimensional metric.
In the non-relativistic limit this approach nicely recovers the 
Lagrangian of covariant Galileons derived in Refs.~\cite{Deffayet}. 
The constructions of more general Galileon theories in the 
framework of branes in a co-dimensional (or maximally symmetric) 
bulk and in supersymmetric theories have been carried 
out in Refs.~\cite{Trodden}. 

Since the field equations of motion following from the action of 
the covariant Galileon are kept up to second order in time and 
spatial derivatives, this theory can avoid the Ostrogradski's 
instability \cite{Ostro} associated with the appearance of 
the Hamiltonian unbounded from below.
The four-dimensional action of the most general scalar-tensor theories 
with second-order equations of motion was first found
by Horndeski in 1974 \cite{Horndeski}.
The same action was re-derived 
by Deffayet {\it et al.} \cite{Deffayet11} 
with a more convenient form in a general $D$-dimensional 
space-time (see also Refs.~\cite{Charmo,Koba11}).
The four-dimensional Horndeski's theory is closely related to 
the effective field theory of inflation \cite{Hoinf} or dark energy \cite{Hoda} 
in that the latter covers the former with extra spatial derivatives 
higher than second order at the level of linear  
cosmological perturbations \cite{Fedo}. 
The Horndeski's theory was applied to the dark energy 
cosmology in Refs.~\cite{Horncosmo,exGAL}.

In the presence of the covariant Galileon Lagrangian 
$M^{-3}X \square \phi$ and non-relativistic matter 
coupled to $\phi$, the Vainshtein mechanism works
to recover the General Relativistic behavior at short distances.
The Vainshtein radius $r_V$ depends on the mass scale $M$.
In the DGP model, $r_V$ can be as large 
as $10^{20}$ cm for $M$ related to dark energy
and the Schwarzschild radius $r_g$ 
of the Sun \cite{DDGV}.
For the distance $r$ satisfying 
$r_g \ll r \ll r_V$ there is the solution 
$\phi'(r) \propto r^{-1/2}$ responsible for the suppression 
of the fifth force within the solar system. 
A similar suppression also occurs for the extended 
Galileon Lagrangian $g(\phi)M^{1-4n}X^n \square \phi$ ($n \geq 1$)
with a non-minimal coupling $F(\phi)R$ \cite{KY,VainARS}, where 
$g(\phi)$ and $F(\phi)$ are slowly varying functions with respect to $\phi$.

In the context of the Horndeski's theory the Vainshtein screening effect  
was studied by Kimura {\it et al.} \cite{Kimura} 
in the spherically symmetric configurations on the cosmological background. 
For the theory in which non-linear field derivatives couple to the 
Einstein tensor (i.e., $G_{5,X} \neq 0$ in the Lagrangian (\ref{L5}) 
given below), Kimura {\it et al.} claimed that 
the Newton gravity is not recovered at short distances. 
It is not clear however that in such models the solution of the field 
in the regime $r_g \ll r \ll r_V$ does not really connect to 
another solution which appears in the region of high density (i.e., inside 
a spherically symmetric body).
In this paper we shall address this issue in detail
by taking into account the variation of the 
matter density inside the body.

In the Horndeski's theory we derive the field equations of motion 
for a spherically symmetric metric characterized by 
two gravitational potentials $\Psi$ and $\Phi$.
Our analysis is general enough to address the Vainshtein 
mechanism for most of modified gravitational 
models proposed in literature (see 
Refs.~\cite{Chow,Burrage,Babichev,BBD,Tana,Wesley,Hiramatsu,Chu,Narikawa,KNT,Bere} 
for the study of the Vainshtein mechanism in related models).
Not only we clarify conditions under which 
the Vainshtein mechanism can be at work, but 
we apply our results to a number of concrete 
models such as covariant/extended Galileons
and the Dirac-Born-Infeld (DBI) Galileons with Gauss-Bonnet 
and other terms.
We obtain the solutions of the scalar field and
the gravitational potentials inside the Vainshtein radius, 
paying particular attention to the 
matching of solutions around the surface of
a spherically symmetric body.

This paper is organized as follows.
In Sec.~\ref{fieldeqsec} the full equations of motion 
are derived in a spherically symmetric space-time
with a matter source. 
On the weak gravitational background we reduce
the equations of the field and gravitational potentials to 
simpler forms.
In Sec.~\ref{vasec} we obtain a general formula of the Vainshtein radius 
in the presence of a non-minimal field coupling $e^{-2Q\phi/M_{\rm pl}}$ 
with the Ricci scalar $R$ and discuss conditions for the existence 
of solutions that accommodate the Vainshtein mechanism.
In Sec.~\ref{galisec} we study in details how the screening 
mechanism operates in the presence of all the covariant Galileon terms.
In Sec.~\ref{appsec} we apply our general results to a number of 
concrete models which are mostly the extension of the covariant 
Galileon. Section \ref{consec} is devoted to conclusions.

\section{Field equations of motion}
\label{fieldeqsec} 

The Lagrangian in the most general scalar-tensor theories 
in four dimensions is described by \cite{Horndeski,Deffayet11,Charmo,Koba11}
\begin{equation}
\mathcal{L}=\sum_{i=2}^{5} \mathcal{L}_{i}\,,\label{Lag}
\end{equation}
where
\begin{eqnarray}
&  & \mathcal{L}_2=K(\phi,X)\,,\label{L2}\\
&  & \mathcal{L}_3=-G_3(\phi,X)\,\Box \phi\,,\label{L3}\\
&  & \mathcal{L}_4=G_4(\phi,X)\, R+G_{4,X}\, \left[(\Box \phi)^2 - (\nabla_{\mu} \nabla_{\nu} \phi)\, (\nabla^{\mu} \nabla^{\nu} \phi)\right] \,,\label{L4}\\
&  & \mathcal{L}_5=G_5(\phi,X) G_{\mu \nu} (\nabla^{\mu} \nabla^{\nu} \phi)-\frac{1}{6}\, G_{5,X} \left[(\Box \phi)^3 -3 (\Box \phi)(\nabla_{\mu} \nabla_{\nu} \phi) (\nabla^{\mu} \nabla^{\nu} \phi)+2  (\nabla^{\mu} \nabla_{\alpha} \phi) (\nabla^{\alpha} \nabla_{\beta} \phi) (\nabla^{\beta} \nabla_{\mu} \phi)\right].\label{L5}
\end{eqnarray}
Here $K(\phi,X)$ and $G_{i} (\phi,X)$ $(i=3,4,5)$ are functions with respect to a scalar field $\phi$ 
and its kinetic energy $X=-g^{\mu \nu}\partial_{\mu}\phi \partial_{\nu} \phi/2$ 
($g^{\mu \nu}$ is the metric tensor), $R$ is the Ricci scalar, and $G_{\mu \nu}$ 
is the Einstein tensor. 
We use the notations $G_{i,X}$ and $G_{i,\phi}$ for the partial 
derivatives of $G_{i}$ with respect to $X$ and $\phi$, respectively. 
Taking into account a barotropic perfect fluid,
the full action is given by 
\begin{equation}
{\cal S}=\int d^{4}x\sqrt{-g}\,{\cal L}
+\int d^4 x\,{\cal L}_m(g_{\mu \nu}, \Psi_m)\,,
\label{action}
\end{equation}
where $g$ is the determinant of the metric $g_{\mu\nu}$, and 
${\cal L}_m$ is the Lagrangian of the matter fields $\Psi_m$.
The energy-momentum tensor of matter is derived from ${\cal L}_m$, 
as $T_{\mu \nu}=-(2/\sqrt{-g})\delta {\cal L}_m/\delta g^{\mu \nu}$.
In terms of the energy density $\rho_m$ and the pressure $P_m$ of
matter, we have that $T^{\mu}_{\nu}={\rm diag}\,(-\rho_{m},P_{m},P_{m},P_{m})$.
We do not introduce the direct coupling between the field $\phi$ 
and matter. In scalar-tensor theories in which the function $G_4$
is a function of $\phi$, the conformal transformation 
to the Einstein frame gives rise to a matter coupling 
with $\phi$ (as we will discuss later).

Let us consider a spherically symmetric space-time
with the distance $r$ from the center of symmetry. 
The line element of such a background is 
\begin{equation}
ds^{2}=-e^{2\Psi(r)}dt^{2}+e^{2\Phi(r)}dr^{2}
+r^{2}(d\theta^{2}+\sin^{2}\theta\, d\phi^{2})\,,\label{line}
\end{equation}
where $\Psi(r)$ and $\Phi(r)$ are functions of $r$. 
On the weak gravitational background 
($|\Psi| \ll 1$, $|\Phi| \ll 1$), the metric (\ref{line}) approximately 
reduces to that of the Newtonian gauge.
For the general metric (\ref{line}) we derive the equations of motion 
valid on the strong gravitational background 
as well\footnote{Recently, Koyama {\it et al.} \cite{KNT} 
expanded the action (\ref{action}) up to second order of perturbations 
around the Minkowski background by imposing that the scalar action 
respects the Galilean symmetry 
$\partial_{\mu} \phi \to \partial_{\mu} \phi+b_{\mu}$. 
In our work we derive the equations of motion from the original 
Horndeski's action without putting any restriction on the functional 
forms of $K$ and $G_i$ ($i=3,4,5$) from the beginning. 
As a result, unlike Ref.~\cite{KNT}, our general equations of motion can 
be used not only for other screening mechanisms such as 
chameleons \cite{chame} and symmetrons \cite{symmetron} 
but also for the study of field configurations on the 
strong gravitational background \cite{stpapers}.}.
The $(00)$, $(11)$ and $(22)$ components of the equations of motion 
following from the action (\ref{action}) are given, 
respectively, by 
\begin{eqnarray}
&  &  \left( A_1+\frac {A_2}{r}+\frac{A_3}{{r}^{2}} \right) \Phi' 
 +A_4+\frac{A_5}{r}+\frac{A_6}{{r}^{2}} = e^{2\Phi} \rho_{m}
 \label{eq:00}\,,\\
 &  &  \left( A_1+\frac {A_2}{r}+\frac{A_3}{{r}^{2}} \right) \Psi' 
 +A_7+\frac{2A_1}{r}+\frac{A_2+2A_8}{2r^{2}} = e^{2\Phi} P_{m}\label{eq:11}\,,\\
 &  &  \left(-e^{-2\Phi}A_8+\frac{A_9}{r}\right) \left(\Psi''+\Psi'^2\right)-\Biggl[ \left(\frac{A_2}{2}+\frac{A_3+e^{2\Phi}A_9}{r} \right) \Psi' +A_1+\frac{A_2}{2r} \Biggr] 
\Phi'  \nonumber \\
& & -\left( \frac{A_5}{2}+\frac{A_6-A_{10}}{r} \right) \Psi' 
-A_4-\frac{A_{5}}{2r}=e^{2\Phi} P_{m}\,,\qquad \label{eq:22}
\end{eqnarray}
where a prime represents the derivative
with respect to $r$.
The coefficients $A_i$ ($i=1,2 \cdots, 10$) are 
\ba
&  &  A_1=-2\,\phi' X{G_{3,X}}+2\,\phi'\,\left({G_{4,\phi}} +2\,X{G_{4,\phi X}}\right)\,, \notag \\
&  &  A_2= 4\,{G_{4}}-16\,X\left({G_{4,X}}+X{G_{4,XX}} \right)+4\,X\left(3\,{G_{5,\phi}}+2\,X{G_{5,\phi X}} \right)\,, \notag \\
&  &  A_3=2\,\phi' \left( 5e^{-2\Phi}-1 \right) X{G_{5,X}}
+4\,\phi' e^{-2\Phi} {X}^{2} {G_{5,XX}}\,, \nonumber\\
&  &  A_4= K e^{2\Phi}-2\, \phi'' \left({G_{4,\phi}}+2\,X {G_{4,\phi X}}\right)
-2e^{2\Phi}X{G_{3,\phi}}+2\,X{G_{3,X}}\, \phi'' 
+4e^{2\Phi}X{G_{4,\phi \phi}}\,,\notag\\
 &  &  A_5=-4\,\phi'\left({G_{4,\phi}-2\,X{G_{4,\phi X}}}\,\right) -4\,\phi'  \phi'' 
 e^{-2\Phi} \left({G_{4,X}}+2\,X{G_{4,XX}}-G_{5,\phi}-X{G_{5,\phi X}} \right) 
 -4\,\phi'X {G_{5,\phi \phi}}\,, \notag \\
 &  &  A_6= -2\,\left( 1-e^{2\Phi} \right) {G_{4}}+4\,X{G_{4,X}}-2\,X\left\{ 
 \left( 1+ e^{2\Phi} \right) {G_{5,\phi}}-2\,X{G_{5,\phi X}}\right\}\notag \\
 &  & \qquad\quad +2\,\phi''X \left\{ \left( 1-3 e^{-2\Phi} \right) {G_{5,X}}
 -2 e^{-2\Phi} X {G_{5,XX}}\right\}\,, \notag \\
 &  &  A_7=-e^{2\Phi}\left(K-2X{K_{,X}}+2XG_{3,\phi} \right)\,, \notag \\
 &  &  A_8= -2 e^{2\Phi} \left( G_4-2XG_{4,X}+XG_{5,\phi} \right)\,,\notag\\
 &  &  A_9=2\,\phi' e^{-2\Phi}{X} {G_{5,X}}\,, \notag\\
 &  &  A_{10}=2\,e^{2\Phi}\left( G_{4}-X{G_{5,\phi}} \right)
 +2 \phi''{X} {G_{5,X}}\,,
\ea
where $X=-e^{-2\Phi}\phi'^{2}/2$.
The matter fluid satisfies the continuity equation 
\begin{equation}
P_{m}'+\Psi'(\rho_{m}+P_{m})=0\,.
\label{continuity}
\end{equation}
Varying the action (\ref{action}) with respect to $\phi$, we obtain 
the equation of motion for the scalar field.
Taking the $r$ derivative of Eq.~(\ref{eq:11}) and substituting 
it into Eq.~(\ref{continuity}) with Eqs.~(\ref{eq:00}) 
and (\ref{eq:22}), we can derive the same field equation 
of motion.

On the weak gravitational background ($|\Phi| \ll 1$ and $|\Psi| \ll 1$) 
the dominant contribution to the l.h.s. of Eq.~(\ref{eq:00}) 
is of the order of $(G_4/r^{2})\Phi$.
For the comparison between each term in Eqs.~(\ref{eq:00}) and (\ref{eq:11}) 
relative to $G_4/r^{2}$, we introduce the following quantities
\begin{eqnarray}
\hspace{-0.5cm}&  &  
\varepsilon_{K}={\frac {e^{2\Phi}K{r}^{2}}{{ 2\,G_4}}}\,,\quad
\varepsilon_{K\phi}=\frac{e^{2\Phi}K_{,\phi}\phi' r^3}{2G_4}\,,\quad
\varepsilon_{KX}=-\frac{e^{2\Phi}XK_{,X}r^2}{G_4}\,,\quad
\varepsilon_{Pm}=\frac{e^{2\Phi}P_m r^2}{2G_4}\,,\quad
\varepsilon_{G3\phi}=-\frac{e^{2\Phi}XG_{3,\phi}r^2}{G_4}\,,
\nonumber \\
\hspace{-0.5cm}& &
\varepsilon_{G3X}=-{\frac {X{ G_{3,X}}\,\phi' r}{{ G_4}}}\,,\quad
\varepsilon_{G4\phi}={\frac {r\phi' { G_{4,\phi}}}{{ G_4}}}\,,\quad
\varepsilon_{G4X}={\frac {2XG_{4,X}}{{ G_4}}}\,,\quad
\varepsilon _{G{5\phi}}={\frac {XG_{5,\phi}}{{2\,G_4}}}\,,
\quad
\varepsilon_{G5X}=\frac{e^{-2\Phi}X G_{5,X}\phi'}{2G_4r}\,,
\label{epdef}
\end{eqnarray}
which are required to be much smaller than 1 to 
recover the General Relativistic behavior inside the solar system.
As long as the Vainshtein mechanism is at work 
one can confirm that $|\varepsilon_{i}| \ll 1$, after deriving the 
solutions to the field equation \cite{VainARS}. 
We also define
\begin{eqnarray}
 &  & \lambda_{K\phi X}={\frac {{K_{,\phi X}}\,\phi' r}{{K_{,X}}}}\,,\quad\lambda_{KXX}={\frac {X{K_{,XX}}}{{K_{,X}}}}\,,\quad\lambda_{G3\phi\phi}={\frac {{G_{3,\phi\phi}}\,\phi' r}{{G_{3,\phi}}}}\,,\quad\lambda_{G3\phi X}={\frac {X{G_{3,\phi X}}}{{G_{3,\phi}}}}\,,\quad\lambda_{G3XX}={\frac {X{G_{3,XX}}}{{G_{3,X}}}}\,,\notag\\
 &  & \lambda_{G4\phi\phi}={\frac {{G_{4,\phi\phi} }\,\phi' r}{{G_{4,\phi} }}}\,,\quad\lambda_{G4\phi X}={\frac {{G_{4,\phi X} }\,\phi' r}{{G_{4,X} }}}\,,\quad\lambda_{G4XX}={\frac {X{G_{4,XX} }}{{G_{4,X} }}}\,,\quad\lambda_{G4\phi\phi X}={\frac {X{G_{4,\phi\phi X}}}{{G_{4,\phi\phi} }}}\,,\notag\\
 &  & \lambda_{G4\phi XX}={\frac {X{G_{4,\phi XX} }}{{G_{4,\phi X} }}}\,,\quad\lambda_{G4XXX}={\frac {X{G_{4,XXX} }}{{G_{4,XX} }}}\,,\quad\lambda_{G5\phi\phi}={\frac {{G_{5,\phi\phi} }\,\phi' r}{{G_{5,\phi}}}}\,,\quad\lambda_{G5\phi X}={\frac {{G_{5,\phi X} }\,\phi' r}{{G_{5,X} }}}\,,\notag\\
 &  & \lambda_{G5XX}={\frac {X{G_{5,XX} }}{{G_{5,X} }}}\,,\quad\lambda_{G5\phi\phi X}={\frac {X{G_{5,\phi\phi X}}}{{G_{5,\phi\phi} }}}\,,\quad\lambda_{G5\phi XX}={\frac {X{G_{5,\phi XX} }}{{G_{5,\phi X} }}}\,,\quad\lambda_{G5XXX}={\frac {X{G_{5,XXX} }}{{G_{5,XX} }}}\,,
\label{lambdadef}
\end{eqnarray}
which are not generally smaller than the order of 1.

For the rest of the paper we use the approximation under which 
all the quantities in Eq.~(\ref{epdef}) are much smaller than 1 
on the weak gravitational background.
{}From Eq.~(\ref{eq:00}) the matter density $\rho_{m}$ is 
of the order of $(G_4/r^{2})\Phi$.
The continuity equation (\ref{continuity}) shows that 
$P_{m}/\rho_{m}\sim\Psi$ in the weak gravitational background
and hence $\varepsilon_{Pm} \sim \Psi^{2}$.
In what follows we neglect the terms coming from the 
gravitational potentials higher than first order (such as $\Psi^2$ and $\Phi^2$)
relative to the parameters $\varepsilon_i$ defined 
in Eq.~(\ref{epdef}). We only keep the first-order 
terms of $\varepsilon_i$. 
We deal with the terms $\varepsilon_i$ multiplied 
by $\lambda_j$ given 
in Eq.~(\ref{lambdadef}) as first-order terms.

Eliminating the terms $\Phi'$ and $\Psi'$ from 
Eqs.~(\ref{eq:00})-(\ref{eq:22}), we obtain
\begin{equation}
\square\Psi=\mu_{1}\rho_{m}+\mu_{2}\square\phi+\mu_{3}\,,
\label{Poisson}
\end{equation}
where $\square \equiv d^2/dr^2+(2/r)(d/dr)$, and 
\begin{eqnarray}
\mu_{1} & \simeq & {\frac {1}{{8G_4}}} \biggl[2+6\,\Phi+\varepsilon_K+\varepsilon_{KX}-\varepsilon_{G3\phi }-\varepsilon_{G3X}-\varepsilon_{G4\phi }- \left( \lambda_{G4\phi X}-2\,\lambda_{G4XX}-3 \right) \varepsilon_{G4X}\,\nonumber \\
&&~~~~~~~~
-8\,\varepsilon_{G5\phi }+ \left( 2\,\lambda_{G5\phi X}-4\,\lambda_{G5XX}-12 \right) \varepsilon_{G5X}\biggr]\,,\label{mu1}\\
\mu_{2} & \simeq & -{\frac {\varepsilon_{G3X}+\varepsilon_{G4\phi }+\lambda_{G4\phi X}\,\varepsilon_{G4X}- 4\,\left( 1+\,\lambda_{G5XX} \right) \varepsilon_{G5X}}{2\,\phi' r}}\,,\label{mu2}\\
\mu_{3} & \simeq & {\frac {2\,\varepsilon_K+\varepsilon_{KX}-\lambda_{G4\phi \phi }\,\varepsilon_{G4\phi }+2\,\lambda_{G4XX}\,\varepsilon_{G4X}+4 \left( \lambda_{G5\phi X}-2\,\lambda_{G5XX}-2 \right) \varepsilon_{G5X}}{2{r}^{2}}}\,.\label{mu3}
\end{eqnarray}

If we rewrite Eq.~(\ref{continuity}) explicitly by using Eqs.~(\ref{eq:00}) and (\ref{eq:11}), 
we find that two Laplacian terms $\square \Psi$ and $\square \phi$ are present. 
Combining this equation with Eq.~(\ref{Poisson}), we can 
eliminate the term $\square \Psi$ to derive the closed-form
equation of $\phi$. 
Using the approximation $e^{2\Phi}\simeq1$, it follows that 
\begin{equation}
\square \phi=\mu_{4}\,\rho_{m}+\mu_{5}\,,
\label{phieq}
\end{equation}
where 
\begin{eqnarray}
\mu_4 & \simeq &  -\frac{r}{4 G_4 \beta}\,\left[ 2\,{G_{4,\phi}}+4\,X{G_{4,\phi X}}-2\,X{G_{3,X}}-\phi' {\beta}-\,{\frac {4\,X \left( {G_{5,X}}+X{G_{5,XX}} \right) }{{r}^{2}}} \right] \,,\label{mu4}\\
 \mu_5  & \simeq & -\frac{1}{r \beta} \biggl[  \left( {K_{,\phi}}-2\,X{K_{,\phi X}}+2\,X{G_{3,\phi \phi}} \right) {r}^{2}-4\,X \left( {K_{,XX}}-2\,{G_{3,\phi X}}+2\,{G_{4,\phi\phi X}} \right) \phi' r-4 X ( 3\,{G_{3,X}}+4\,X{G_{3,XX}}  \nonumber \\
& &~~~~~~~~-9{G_{4,\phi X}}-10\,X{G_{4,\phi XX}} 
 +X{G_{5,\phi\phi X}} )+\,{\frac {8X\phi'  \left(3\,{G_{4,XX}}+ 2\,X{G_{4,XXX}}-2\,{G_{5,\phi X}} -X{G_{5,\phi XX}}\right) }{r}} \biggr]\,, \label{mu5} \\
\beta &\equiv&  \left({K_{,X}}+ 2\,X{K_{,XX}}-2\,{G_{3,\phi}}-2\,X{G_{3,\phi X}} \right) r-4\,\phi'  \left( {G_{3,X}}+X{G_{3,XX}}-3\,{G_{4,\phi X}}-2\,X{G_{4,\phi XX}} \right)  \nonumber \\
 &  & -{\frac {4\,X \left( 3\,{G_{4,XX}}+2\,X{G_{4,XXX}}-2\,{G_{5,\phi X}}-X{G_{5,\phi XX}} \right) }{r}}\,.\label{beta}
\end{eqnarray}
After the linear expansion with respect to the parameters $\varepsilon_i$,
we reverted to use the original functions $K$ and $G_i$.

Eliminating the term $\square \phi$ from Eqs.~(\ref{Poisson}) and 
(\ref{phieq}), we obtain the modified Poisson equation
\begin{equation}
\square\Psi=4\pi G_{{\rm eff}}\rho_{m}+\mu_{2}\mu_{5}+\mu_{3}\,,
\label{Poimo}
\end{equation}
where 
\begin{eqnarray}
G_{{\rm eff}} & \equiv & \frac{1}{4\pi}(\mu_{2}\mu_{4}+\mu_{1})\nonumber \\
&= & \frac{1}{16 \pi G_4} 
\left[ 1+\frac{r}{G_4 \beta} \alpha \left( \alpha -\frac12 \phi' \beta \right)
+3\Phi+{\cal O}(\varepsilon_i) \right]\,, \\
\alpha & \equiv & G_{4,\phi}+2XG_{4,\phi X}-XG_{3,X}
-\frac{2X(G_{5,X}+XG_{5,XX})}{r^2}\,.
\label{Geff}
\end{eqnarray}

In GR where the functions are given by 
$G_4=M_{\rm pl}^2/2$, $G_3=0=G_5$ ($M_{\rm pl}$ is 
the reduced Planck mass related to the gravitational constant $G$, 
as $M_{\rm pl}=(8\pi G)^{-1}$), it follows that 
$\alpha=0$, $G_{\rm eff}=G[1+3\Phi+{\cal O}(\varepsilon_i)]$, 
$\mu_2=0$, and $\mu_3={\cal O}(\varepsilon_i)/r^2$.
As long as $|\varepsilon_i| \ll \{ |\Phi|, |\Psi| \}$ (which is usually 
the case for a scalar field responsible for dark energy), 
the gravitational potentials are not affected by the presence 
of the field $\phi$.

The modification of gravity manifests itself for the theories 
characterized by 
\begin{equation}
\alpha \neq 0\,.
\end{equation}
The representative example having a non-zero value of 
$\alpha$ is the dilatonic coupling \cite{Gasperini} given by 
\begin{equation}
G_4 (\phi)=\frac{M_{\rm pl}^2}{2} e^{-2Q\phi/M_{\rm pl}}\,,
\label{G4cho}
\end{equation}
where $Q$ is a coupling constant of the order of unity.
If we consider a canonical massless field,
i.e., $K=X$ and $G_3=0=G_5$, we have that 
$G_{\rm eff} \simeq G \{ 1+2Q^2 [1+\phi'r/(2QM_{\rm pl})]
+3\Phi+{\cal O}(\varepsilon_i) \}$ in the regime 
$|\phi/M_{\rm pl}| \ll 1$.
For $|Q|$ of the order of 1, the deviation of $G_{\rm eff}$ from 
$G$ is significant due to the presence of the term $2Q^2$.
In such cases we need to resort to some mechanism to 
suppress the propagation of the fifth force.
 
Provided that the term $\mu_2 \square \phi$ in Eq.~(\ref{Poisson}) 
is suppressed relative to other terms, the General Relativistic 
behavior can be recovered at short distances.
There are several known mechanisms to screen the fifth
force\footnote{There is also another mechanism called 
the runaway dilaton scenario \cite{runaway}. 
In this model the dilatonic coupling is assumed to be
$G_4=M_{\rm pl}^2/2+B e^{-\mu \phi/M_{\rm pl}}$, where $B$ 
and $\mu~(>0)$ are constants. As the dilaton runs away toward 
the regime $\phi \gg M_{\rm pl}$, $G_4$ approaches 
the value $M_{\rm pl}^2/2$ to recover the General Relativistic behavior.}:
(i) the chameleon mechanism \cite{chame}, 
(ii) the symmetron mechanism \cite{symmetron}, and 
(iii) the Vainshtein mechanism \cite{Vainshtein}.
All of them are covered in our general set-up.

Both the chameleon and symmetron mechanisms are based on 
the presence of the field potential $V(\phi)$. 
To be more concrete, let us consider the theories described by the action 
\begin{equation}
{\cal S}=\int d^4 x \sqrt{-g} \left[ \frac{M_{\rm pl}^2}{2}
F(\phi)R+\omega(\phi)X-V(\phi) \right]
+\int d^4 x\,{\cal L}_m(g_{\mu \nu}, \Psi_m)\,,
\label{actionpo}
\end{equation}
where $F(\phi)$ and $\omega(\phi)$ are functions of $\phi$.
The Brans-Dicke theory \cite{Brans} with the potential $V(\phi)$
corresponds to $F(\phi)=e^{-2Q \phi/M_{\rm pl}}$ and 
$\omega(\phi)=(1-6Q^2)e^{-2Q\phi/M_{\rm pl}}$ \cite{Yoko}.
The coupling $Q$ is related to the Brans-Dicke parameter
$\omega_{\rm BD}$, as $3+2\omega_{\rm BD}=1/(2Q^2)$ \cite{chame,Yoko}.
The metric $f(R)$ gravity and the dilaton gravity are the 
sub-class of Brans-Dicke theory with the parameters
$\omega_{\rm BD}=0$ ($Q^2=1/6$) \cite{Ohanlon} and 
$\omega_{\rm BD}=-1$ ($Q^2=1/2$) \cite{Gasperini}, 
respectively. In this case the field equation of motion 
(\ref{phieq}) reads 
\begin{equation}
\square \phi=\frac{1}{\omega} \left(
-\frac{M_{\rm pl}^2 F_{,\phi}-\phi' \omega r}{2M_{\rm pl}^2 F}
\rho_m+X\omega_{,\phi}+V_{,\phi} \right)\,.
\label{squarephi}
\end{equation}
The coupling such as $F(\phi)=e^{-2Q \phi/M_{\rm pl}}$
gives rise to a matter coupling term $Q\rho_m/M_{\rm pl}$
inside the parenthesis of Eq.~(\ref{squarephi}).
In order to have the description of a canonical field coupled
to matter, it is convenient to transform the action (\ref{actionpo})
to that in the Einstein frame by a conformal 
transformation $\hat{g}_{\mu \nu}=F(\phi)g_{\mu \nu}$ \cite{Maeda}.
The action in the Einstein frame is given by 
\begin{equation}
\hat{{\cal S}}=\int d^4 x \sqrt{-\hat{g}} \left[ \frac{M_{\rm pl}^2}{2}
\hat{R}-\frac12 \hat{g}^{\mu \nu} \partial_{\mu}\chi
\partial_{\nu}\chi-\hat{V}(\chi) \right]+
\int d^4 x\,{\cal L}_m(A^2(\chi)\hat{g}_{\mu \nu}, \Psi_m)\,,
\label{Einaction}
\end{equation}
where
\begin{equation}
\chi \equiv \int d\phi\,\sqrt{\frac32 \left(\frac{M_{\rm pl}F_{,\phi}}{F} \right)^2+
\frac{\omega}{F}}\,,\qquad
\hat{V}(\chi) \equiv \frac{V}{F^2}\,,\qquad
A^2 (\chi) \equiv F^{-1}(\phi)\,.
\end{equation}

Variation of the action (\ref{Einaction}) with respect to the canonical field 
$\chi$ gives $\hat{\square}\chi=\hat{V}_{,\chi}-(1/\sqrt{-\hat{g}})
(\partial {\cal L}_m/\partial \chi)$.
The field $\chi$ couples to matter, as 
$\partial {\cal L}_m/\partial \chi=(A_{,\chi}/A)\sqrt{-\hat{g}}\,\hat{T}$, 
where $\hat{T}=-\hat{\rho}_m+3\hat{P}_m \simeq -\hat{\rho}_m$ 
is the trace of non-relativistic matter in the Einstein frame.
The energy density $\rho_m$ in the Jordan frame is related to $\hat{\rho}_m$
via $\hat{\rho}_m=A^4 \rho_m$.
Using the conserved energy density 
$\rho_m^{*}=A^3 \rho_m=\hat{\rho}_m/A$ in the Einstein frame, 
the field equation reads \cite{chame}
\begin{equation}
\hat{\square}\chi=\hat{V}_{{\rm eff},\chi}\,,\qquad
\hat{V}_{\rm eff} (\chi) \equiv \hat{V} (\chi)+A(\chi) \rho_m^*\,.  
\label{chiein}
\end{equation}

In Brans-Dicke theory with the functions $F(\phi)=e^{-2Q \phi/M_{\rm pl}}$ and 
$\omega(\phi)=(1-6Q^2)e^{-2Q\phi/M_{\rm pl}}$, the field $\chi$ is 
equivalent to $\phi$ and hence $A(\chi)=e^{Q \chi/M_{\rm pl}}$. 
For a runaway potential $\hat{V}(\chi)$, the effective potential $\hat{V}_{\rm eff}(\chi)$
can have a minimum at $\hat{V}_{{\rm eff},\chi} (\chi_{M})=0$ 
due to the presence of the matter coupling $e^{Q \chi/M_{\rm pl}}\rho_m^*$.
The effective mass $m_{\chi}$ of the field at $\chi=\chi_{M}$
depends on the matter density $\rho_m^*$.
Provided that the mass $m_{\chi}$ is large in the region of high density and 
that a spherically symmetric body has a thin shell around its surface, 
the propagation of the fifth force is suppressed
outside the body. This is the screening effect of the chameleon 
mechanism. The solution of Eq.~(\ref{chiein}) and the resulting 
local gravity constraints on concrete potentials [such 
as $V(\phi)=M^{4+n}\phi^{-n}$ and $V(\phi)=M^4 \exp(M^n/\phi^n)]$
were studied in detail in Refs~\cite{chame,DMota}, so we do not repeat them here.
The application of the chameleon mechanism to dark energy 
models based on $f(R)$ gravity and Brans-Dicke theory was carried out in 
Refs.~\cite{fRlocal,Yoko}.

In the symmetron mechanism the choices of the coupling $A(\chi)$ 
and the potential $\hat{V}(\chi)$ are different from those in the chameleon 
mechanism. They are given by \cite{symmetron}
\begin{equation}
A(\chi)=1+\frac{\chi^2}{2M^2}\,,\qquad
\hat{V}(\chi)=-\frac12 \mu^2 \chi^2+\frac14 \lambda \chi^4\,,
\label{sympo}
\end{equation}
where $M$, $\mu$, $\lambda$ are constants.
In this case the effective potential (\ref{chiein}) reads 
\begin{equation}
\hat{V}_{\rm eff} (\chi)=\frac12 \left( \frac{\rho_m^{*}}{M^2}
-\mu^2 \right) \chi^2+\frac14 \lambda \chi^4\,,
\end{equation}
up to an irrelevant constant.
For large $\rho_m^*$ the effective potential is 
$\hat{V}_{\rm eff} \simeq \rho_m^{*}\chi^2/(2M^2)$ 
and hence the field is nearly frozen around $\chi=0$.
For small $\rho_m^*$ the $Z_2$ symmetry is spontaneously 
broken, so that the field has a vacuum expectation value 
$\chi_0=\mu/\sqrt{\lambda}$ in the limit $\rho_m^* \to 0$. 
Since the propagation of the fifth force is suppressed 
in the region of high density, it is possible for the symmetron 
field to pass local gravity constraints.
The experimental bounds and the cosmological implication 
of symmetrons were studied in detail 
in Refs.~\cite{symap}.

In both the chameleon and symmetron mechanisms the
$\hat{\square} \chi$ term in Eq.~(\ref{chiein}) is suppressed in 
the regions of high density, in which case the field-dependent term 
$\mu_2 \square \phi$
in Eq.~(\ref{Poisson}) is effectively decoupled from gravity.
The viability of these two mechanisms heavily depends on 
the choice of the field potentials.
If the chameleon field is responsible for the cosmic acceleration today, 
the potential needs to be carefully designed to satisfy both 
cosmological and local gravity 
constraints \cite{Yoko}.
It is also known that the energy scale of the simplest symmetron 
potential (\ref{sympo}) is too small to be used 
for dark energy \cite{symmetron}.

The Vainshtein mechanism, which is the main topic of our paper, 
is based on non-linear field self-interactions like $X \square \phi$.
In such cases, the last two terms inside the parentheses 
on the r.h.s. of Eqs.~(\ref{mu5}) and (\ref{beta}) provide the dominant contributions
for the distance smaller than the so-called Vainshtein radius $r_V$.
In this regime we have $|\mu_4 \rho_m| \ll |\mu_5|$ and hence
$\square \phi \simeq \mu_5$ \cite{VainARS}. 
For the choice $G_3 \propto X$ and the coupling (\ref{G4cho})
the solution to the field equation 
is given by $\phi'(r) \propto r^{-1/2}$, so that 
the terms $\mu_2 \mu_4$ and $\mu_2 \mu_5$ in Eq.~(\ref{Poimo})
are suppressed relative to other terms.

In this paper we study the effects of other non-linear field 
self-interactions such as those coming from $G_4(\phi, X)$
and $G_5(\phi,X)$ in addition to the term $G_3(\phi,X)$.
For concreteness we choose the field coupling of 
the form (\ref{G4cho}) and non-linear field derivative couplings. 
The coupling (\ref{G4cho}) is sufficiently 
general in that it covers a wide variety of theories such as
Brans-Dicke theory and dilaton gravity.
Moreover it generally appears after the dimensional reduction 
in higher-dimensional theories (the field $\phi$ characterizes 
the size of compact space or the position 
of a probe brane) \cite{Davis,Vainother,Rham}.

Unlike the chameleon and symmetron scenarios, the Vainshtein 
mechanism can be at work even without the field potential $V(\phi)$.
We adopt the k-essence Lagrangian of the form 
$K(\phi,X)=f_2(\phi)g_2(X)$ \cite{Mukhanov} without an explicit 
potential $V(\phi)$, 
where the function $g_2(X)$ includes the non-linear terms of $X$.
For the functions $G_i(\phi,X)$ ($i=3,4,5$) we also consider the 
couplings of the form $f_i(\phi) g_i(X)$.
In summary we focus on the theories characterized by 
\begin{eqnarray}
& &K(\phi,X)=f_2(\phi)g_2(X)\,,\qquad
G_3(\phi,X)=f_3(\phi)g_3(X)\,,\nonumber \\
& &G_4(\phi,X)=\frac{M_{\rm pl}^2}{2} e^{-2Q\phi/M_{\rm pl}}
+f_4(\phi)g_4(X)\,,\qquad
G_5(\phi,X)=f_5(\phi)g_5(X)\,.
\label{gencho}
\end{eqnarray}
For concreteness we take the exponential couplings of the form
\begin{eqnarray}
f_i(\phi)=e^{-\lambda_i \phi/M_{\rm pl}}\,,\qquad (i=2,3,4,5),
\label{fphi}
\end{eqnarray}
where $\lambda_i$'s are constants. 
The choice of (\ref{fphi}) is motivated by the dilatonic couplings 
appearing in low-energy effective string theory. 
The constants $\lambda_i$ and $Q$ are assumed to be 
at most of the order of unity.
Provided that the Vainshtein mechanism is at work, the field can 
stay in the regime $|\phi/M_{\rm pl}| \ll 1$.
In most cases the models with constant $f_i(\phi)$ do not exhibit 
significant differences from those with the exponential couplings (\ref{fphi}), 
but there are some specific models in which the presence 
of the field-dependent couplings can change the behavior 
of solutions (such as those discussed in Sec.~\ref{secnonzeroc5}).
In Secs.~\ref{galisec} and \ref{appsec} we clarify this issue in detail.

The non-linear self-interaction $g_3(X)$ proportional to $X$ arises 
in the DGP braneworld scenario \cite{DGP,DGPnon} and 
in the Kaluza-Klein theory with a higher-dimensional 
Gauss-Bonnet term \cite{Vainother}.
The covariant Galileon \cite{Deffayet}, whose Lagrangian arises 
as a non-relativistic limit for a probe brane embedded in a five-dimensional 
bulk \cite{Rham}, corresponds to the choice $g_2(X)=X$, 
$g_3(X)=X$, $g_4(X)=X^2$, and $g_5(X)=X^2$, with 
constant $f_i(\phi)$ (i.e., $\lambda_i=0$).
The extended Galileon \cite{KY,exGAL} has more general powers $p_i$
of the derivative terms, i.e., $g_{i}(X)=X^{p_i}$.
The four-dimensional Gauss-Bonnet coupling $-f(\phi)R_{\rm GB}^2$ 
can be also 
accommodated in Eq.~(\ref{gencho}) for 
specific choices of $f_i(\phi)$ and $g_i(X)$ \cite{Koba11}.

\section{Vainshtein mechanism}
\label{vasec} 

Let us study how the Vainshtein mechanism generally works for the theories 
given by the functions (\ref{gencho}). We assume that the coupling 
$Q$ is of the order of unity.
The field non-linear self-interactions $f_i(\phi) g_i(X)$ in 
$G_i(\phi,X)$ ($i=3,4,5$) can be 
responsible for the suppression of the fifth force within the 
so-called Vainshtein radius $r_V$.
In the following we study general solutions of the field equation 
of motion (\ref{phieq}) in the regimes (A) $r \gg r_V$, 
(B) $r_g \ll r \ll r_V$, and (C) $r < r_s$, separately, 
where $r_g$ is the Schwarzschild radius of a star 
with the radius $r_s$.
In Sec.~\ref{concretemodel} we apply our results to a concrete model to discuss
the matching of solutions in three different regimes.
In the same model we also derive the explicit expression of
the gravitational potentials inside the Vainshtein radius.

\subsection{$r \gg r_V$}
\label{rlarge}

For the distance $r$ much larger than $r_V$, the non-linear field-self
interactions are suppressed in Eqs.~(\ref{mu4})-(\ref{beta}).
In Eq.~(\ref{mu4}) this means that the term $2G_{4,\phi}$ is the 
dominant contribution, i.e., 
\be
|2G_{4,\phi}| \gg |4X G_{4,\phi X}-2XG_{3,X}-\phi' \beta
-4X (G_{5,X}+XG_{5,XX})/r^2|\,.
\label{con1la}
\ee
For the function $G_4(\phi,X)$ given in Eq.~(\ref{gencho}), 
the following condition should be satisfied
\be
M_{\rm pl}^2 e^{-2Q \phi/M_{\rm pl}}/2 \gg 
e^{-\lambda_4 \phi/M_{\rm pl}}g_4(X)\,.
\label{con2la}
\ee
Moreover, we assume that the field is in the range
\be
|\phi/M_{\rm pl}| \ll 1\,,
\label{con3la}
\ee
which can be justified after deriving the solution 
to Eq.~(\ref{phieq}).

The function $g_2(X)$ inside $K(\phi, X)$ may be written 
in terms of the sum of the polynomials, as
$g_2(X)=\sum_{n=1}^{\infty} c_{n}\, \mu^4 \left( X/\mu^4
\right)^n$, 
where $c_{n}$'s are dimensionless constants and $\mu$ is 
another constant having a dimension of mass.
In the following we focus on the model in which the first term 
in $g_2(X)$ dominates over the other terms, i.e., 
$g_2(X) \simeq c_1 X$. Without loss of generality we can 
choose the coefficient to be $c_1=1$, so that the function 
$K(\phi,X)$ is 
\begin{equation}
K(\phi,X)=e^{-\lambda_2 \phi/M_{\rm pl}}X\,.
\label{kba}
\end{equation}
Since the term $K_{,X}\,r$ should be
the dominant contribution in Eq.~(\ref{beta}), we have
\ba
r &\gg &| 2(G_{3,\phi}+XG_{3,\phi X})r+4(G_{3,X}+XG_{3,XX}
-3G_{4,\phi X}-2XG_{4,\phi XX}) \phi' \nonumber \\
& &-2(3G_{4,XX}+2XG_{4,XXX}
-2G_{5, \phi X}-XG_{5,\phi XX})\phi'^2/r|.
\label{con4la}
\ea
We are also in the regime where the matter-coupling term 
$\mu_4 \rho_m$ dominates over another term $\mu_5$, i.e.,
\begin{equation}
|\mu_4| \rho_m \gg |\mu_5|\,.
\label{con5la}
\end{equation}

Since $\mu_4 \simeq Q/M_{\rm pl}$ under the above conditions, 
Eq.~(\ref{phieq}) reads
\begin{equation}
\frac{d}{dr} (r^2 \phi') \simeq QM_{\rm pl}
\frac{dr_g}{dr}\,,
\label{dreq1}
\end{equation}
where the Schwarzschild radius $r_g$ is defined by 
\begin{equation}
r_g \equiv \frac{1}{M_{\rm pl}^2} \int_0^{r}
\rho_m \tilde{r}^2 d\tilde{r}\,.
\label{rgdef}
\end{equation}
Integration of Eq.~(\ref{dreq1}) gives the following solution 
\begin{equation}
\phi' (r) = \frac{QM_{\rm pl} r_g}{r^2} \qquad
(r \gg r_V).
\label{phila}
\end{equation}
This gives rise to the fifth force of the order of $|\phi'(r)/M_{\rm pl}|=|Q|r_g/r^2$, 
by which the gravitational law is significantly modified.
On using the boundary condition $\phi( \infty) \to 0$, 
we obtain
\begin{equation}
\phi (r) = -\frac{QM_{\rm pl} r_g}{r}\,.
\end{equation}
Then, the condition (\ref{con3la}) translates into $r \gg r_g$
for $|Q|={\cal O}(1)$. 
Since we are now in the regime $r \gg r_V$, it can be 
interpreted as
\begin{equation}
r_V \gg r_g\,.
\label{rVg}
\end{equation}
For a given model (i.e., for given functions of $G_{3,4,5}$)
we need to confirm whether the conditions 
(\ref{con1la}), (\ref{con2la}), (\ref{con4la}), and 
(\ref{con5la}) are satisfied.
In Sec.~\ref{concretemodel} we confirm those conditions 
for a concrete model.

\subsection{$r_g \ll r \ll r_V$}
\label{rsmall}

For the distance much smaller than $r_V$, the non-linear field 
self-interactions are the dominant contribution in Eq.~(\ref{phieq}).
The Vainshtein radius is characterized by the distance at which 
the field self-interactions become comparable to the term 
$K_{,X}\,r$, i.e., 
\ba
r_V &=&| 2(G_{3,\phi}+XG_{3,\phi X})r_V+4(G_{3,X}+XG_{3,XX}
-3G_{4,\phi X}-2XG_{4,\phi XX}) \phi' (r_V) \nonumber \\
& &-2(3G_{4,XX}+2XG_{4,XXX}-2G_{5, \phi X}-XG_{5,\phi XX})
\phi'^2(r_V)/r_V|.
\label{varadius}
\ea
For a given model, the Vainshtein radius is explicitly known 
by employing the solution (\ref{phila}).
The regime $r \ll r_V$ corresponds to the opposite inequality 
of Eq.~(\ref{con4la}), i.e.,
\ba
r &\ll &| 2(G_{3,\phi}+XG_{3,\phi X})r+4(G_{3,X}+XG_{3,XX}
-3G_{4,\phi X}-2XG_{4,\phi XX}) \phi' \nonumber \\
& &-2(3G_{4,XX}+2XG_{4,XXX}
-2G_{5, \phi X}-XG_{5,\phi XX})\phi'^2/r|.
\label{con1sm}
\ea
The terms inside Eq.~(\ref{mu5}) should satisfy the following condition 
\ba
|(\lambda_2/M_{\rm pl}+2 G_{3,\phi \phi}) r^2
+4(2G_{3,\phi X}-2G_{4,\phi \phi X}) \phi' r| 
&\ll& |4 (3G_{3,X}+4XG_{3,XX}-9G_{4,\phi X}-10X G_{4,\phi XX}
+X G_{5,\phi \phi X})  \nonumber \\
& &-8\phi'(3G_{4,XX}+2XG_{4,XXX}-2G_{5,\phi X}-XG_{5,\phi XX})/r|\,.
\label{con2sm}
\ea
As long as the Vainshtein mechanism is at work, the matter coupling 
term $\mu_4 \rho_m$ should be suppressed relative to the term 
$\mu_5$, i.e., 
\be
|\mu_4| \rho_m \ll |\mu_5|\,.
\label{con3sm}
\ee

Under the conditions (\ref{con1sm})-(\ref{con3sm})
the field equation (\ref{phieq}) reads
\be
\frac{d}{dr} (r^2 \phi') \simeq \frac{\xi_1}{\xi_2} r \phi' \,,
\label{fieldap}
\ee
where 
\ba
\xi_1&\equiv& r(3G_{3,X}+4XG_{3,XX}
-9G_{4,\phi X}-10XG_{4, \phi XX}+XG_{5,\phi \phi X}) \nonumber \\
& &-2\phi' (3G_{4,XX}+2XG_{4,XXX}-2G_{5,\phi X}-XG_{5,\phi XX})\,,
 \nonumber \\
\xi_2 &\equiv& 2r (G_{3,X}+XG_{3,XX}-3G_{4, \phi X}
-2X G_{4, \phi XX})-\phi'
(3G_{4,XX}+2XG_{4,XXX}-2G_{5,\phi X}-XG_{5,\phi XX})\,.
\label{xi1}
\ea
For a given model, the solution to $\phi'(r)$ is 
known by integrating Eq.~(\ref{fieldap}).
After deriving the solution, we need to confirm whether the conditions 
(\ref{con1sm})-(\ref{con3sm}) are satisfied in the regime
$r \ll r_V$. In Sec.~\ref{concretemodel} we study the solution of 
Eq.~(\ref{fieldap}) for a concrete model to understand the 
consistency of the conditions used above.
For the validity of the solution we typically require 
that the distance $r$ is much larger than $r_g$ \cite{VainARS}. 
This is related to the fact that inside a spherically symmetric body
the matter density $\rho_m$ becomes large, so that the condition 
(\ref{con3sm}) tends to be violated.

\subsection{$r<r_s$}
\label{rcenter}

If the condition (\ref{con3sm}) is violated inside a star 
with the radius $r_s$, we can no longer use the solution 
to the field equation (\ref{fieldap}). 
At the origin we generally impose the following 
boundary condition 
\be
\phi'(0)=0\,.
\label{phicen}
\ee
If all the non-linear terms $f_i(\phi)g_i(X)$ 
in Eq.~(\ref{gencho}) are suppressed relative to the term 
$G_4=M_{\rm pl}^2e^{-2Q\phi/M_{\rm pl}}/2$, the 
field equation (\ref{phieq}) reduces to the same form as 
Eq.~(\ref{dreq1}), i.e.,
$d(r^2\phi')/dr \simeq Q\rho_m r^2/M_{\rm pl}$. 
Assuming that $\rho_m$ is nearly constant inside the body, 
we obtain the integrated solution
$\phi'(r) \simeq Q\rho_m r/(3M_{\rm pl})$. 
In fact, this satisfies the boundary condition (\ref{phicen}).
However, if the solution inside the body corresponds to 
this type, it gives rise to a large modification of gravity 
around the surface of the star because it is the analogue
of (\ref{phila}) outside the star.

In the presence of the non-linear field self-interactions, 
the solutions to the field equation (\ref{fieldap}) inside the body 
are different from $\phi'(r) \simeq Q\rho_m r/(3M_{\rm pl})$. 
As we will see in the following sections, the solutions 
depend on the choice of the functions $G_i(\phi,X)$.
As long as the Vainshtein mechanism operates both 
inside and outside the star, we will show that the interior and 
exterior solutions smoothly connect each other. 
In order to study the matching of the solutions properly, 
we need to assume the density profile of the star.
In the numerical simulations given in the following sections, 
we employ the profile
\be
\rho_m=\rho_c \exp \left(-r^2/r_t^2 \right)\,,
\label{profile}
\ee
where $\rho_c$ is the central density of 
the body with the radius $r_s$.
The density $\rho_m$ starts to decrease significantly 
around the distance $r_t$. 
We confirmed that the different choices of the density profile 
do not affect our main results.

\subsection{Concrete example}
\label{concretemodel}

Let us consider the covariant Galileon model \cite{Deffayet}
in the presence of the term $G_4$ alone, i.e., 
\be
G_4 (\phi,X)=\frac{M_{\rm pl}^2}{2} e^{-2Q\phi/M_{\rm pl}}+
\frac{c_4}{M^6}X^2\,,\qquad
G_3=G_5=0\,,
\ee
where $c_4$ is a dimensionless constant of the order of 1, 
and $M$ is another constant having the dimension of mass.
Substituting the solution (\ref{phila}) into Eq.~(\ref{varadius}), 
we obtain the Vainshtein radius
\be
r_V=(12|c_4|)^{1/6} \frac{(|Q|M_{\rm pl}r_g)^{1/3}}{M}
\approx \frac{(|Q|M_{\rm pl}r_g)^{1/3}}{M}\,.
\label{rV1}
\ee

The field equation (\ref{fieldap}) reduces to
\be
\frac{d}{dr} (r^2 \phi')=2r\phi'\,.
\ee
The solution to this equation is simply given by 
\be
\phi'(r)=C\,,
\label{phiin}
\ee
where $C$ is a constant. 
Matching (\ref{phiin}) with (\ref{phila}) at $r=r_V$, 
it follows that 
\be
\phi'(r)=\frac{QM_{\rm pl}r_g}{r_V^2} \qquad
(r_g \ll r \ll r_V).
\label{phiin2}
\ee
Using the boundary condition $\phi(r_V)=-QM_{\rm pl}r_g/r_V$, 
we obtain the following solution 
\be
\phi(r)=\frac{QM_{\rm pl} r_g}{r_V}
\left( \frac{r}{r_V}-2 \right)\,.
\ee
This shows that even in the regime $r_g \ll r \ll r_V$
the condition (\ref{con3la}) is satisfied for $r_V \gg r_g$.
Now we check the consistency of several other conditions 
used in Secs.~\ref{rlarge} and \ref{rsmall}.

Let us first consider the regime $r \gg r_V$
with the solution (\ref{phila}).
The condition (\ref{con4la}) simply corresponds to
$r \gg r_V$, where $r_V$ is given by Eq.~(\ref{rV1}).
Since $\beta \simeq r$ in this case
the condition (\ref{con1la}) translates to $r \gg r_g$, 
which is equivalent to (\ref{rVg}).
The condition (\ref{con2la}) reduces to 
$(r/r_V)^8 \gg Q^2 (r_g/r_V)^2/24$, which is 
automatically satisfied for $r_V \gg r_g$.
Since $\mu_4 \simeq Q/M_{\rm pl}$ and 
$\mu_5 \simeq 24c_4 Q^3M_{\rm pl}^3r_g^3/(M^6 r^9)$
for $r$ not away from $r_V$, the condition (\ref{con5la})
corresponds to 
\be
r \gg r_* \equiv \left( 24|c_4|Q^2 \frac{M_{\rm pl}^4 r_g^3}
{M^6 \rho_m} \right)^{1/9} \approx 
\left( \frac{Q^2 M_{\rm pl}^4 r_g^3}
{M^6 \rho_m} \right)^{1/9}\,.
\label{rstar}
\ee
The critical radius $r_*$ depends on the mass scale $M$
and the density profile $\rho_m$.
The ratio between $r_*$ and $r_V$ is given by 
\be
\frac{r_*}{r_V} \approx \left( \frac{M^3 M_{\rm pl}}
{|Q| \rho_m} \right)^{1/9}\,.
\label{rratio}
\ee
If the same model is responsible for the late-time cosmic acceleration, 
the mass scale $M$ is related to the today's Hubble parameter $H_0$, 
as $M^3 \approx M_{\rm pl}H_0^2$ \cite{DTPRL}.
Since the critical density $\rho_0 \approx 10^{-29}$ g/cm$^3$
has the relation $\rho_0 \approx M_{\rm pl}^2 H_0^2$, 
the ratio (\ref{rratio}) can be estimated as 
$r_*/r_V \approx (\rho_0/\rho_m)^{1/9}$ for $|Q|={\cal O}(1)$.
If $\rho_m$ is close to $\rho_0$, then $r_* \approx r_V$.
The Schwarzschild radius of the Sun is $r_g \simeq 3 \times 10^{5}$~cm, 
in which case the Vainshtein radius can be estimated as 
$r_V \approx (r_g H_0^{-2})^{1/3} \approx 10^{20}$~cm for 
$M^3=M_{\rm pl}H_0^2$.
This radius is much larger than the solar-system scale.
Even by taking the mean density $\rho_m \approx 10^{-24}$~g/cm$^3$
of our galaxy, $r_*$ is the same order as $r_V$.
The above discussion shows that all the conditions 
(\ref{con1la}), (\ref{con2la}), (\ref{con4la}), and (\ref{con5la}) are 
satisfied in the regime $r \gg r_V$.

We proceed to the regime $r_g \ll r \ll r_V$ characterized by the solution (\ref{phiin2}).
The condition (\ref{con1sm}) exactly corresponds to $r \ll r_V$.
For $|\lambda_2|={\cal O}(1)$ the condition (\ref{con2sm}) translates 
to $(r/r_V)^3 \ll r_V/r_g$, which is automatically satisfied for 
$r_V \gg r_g$. In the regime $r \gg r_g$
we have that $\mu_4 \simeq Q(r/r_V)^2/(12c_4 M_{\rm pl})$
and $\mu_5 \simeq 2QM_{\rm pl} r_g/(r_V^2 r)$. 
Then, the condition (\ref{con3sm}) can be interpreted as 
$r \ll \tilde{r}_* \equiv (M_{\rm pl}^2 r_g/\rho_m)^{1/3}$.
Since $\tilde{r}_*/r_V \approx (M^3 M_{\rm pl}/\rho_m)^{1/3}$, 
$\tilde{r}_*$ is close to $r_V$ for $M^2 \approx M_{\rm pl}H_0^2$
and $\rho_m \approx \rho_0$.
Thus, all the conditions used to derive the solution 
(\ref{phiin2}) are consistently satisfied.
We can also check that the quantities $\varepsilon_i$ defined 
in Eq.~(\ref{epdef}) remain much smaller than the order of 1.

Picking up the dominant terms of Eqs.~(\ref{eq:00}) 
and (\ref{eq:11}) in the regime 
$r_g \ll r \ll r_V$, it follows that 
\ba
\frac{d}{dr} (r \Phi) &\simeq&
-\frac{2Q\phi' r}{M_{\rm pl}}
+\frac{\rho_m r^2}{2M_{\rm pl}^2}\,,
\label{Phieq} \\
\Psi' & \simeq& \frac{\Phi}{r}+\frac{2Q\phi'}{M_{\rm pl}}\,.
\label{Psieq}
\ea
Using (\ref{phiin2}), we obtain the following integrated solutions
\be
\Phi \simeq \frac{r_g}{2r} \left[ 1-2Q^2 \left(
\frac{r}{r_V} \right)^2 \right]\,, \qquad
\Psi \simeq -\frac{r_g}{2r} \left[ 1-2Q^2 \left(
\frac{r}{r_V} \right)^2 \right]\,.\label{Psiso1}
\ee
We define the post-Newtonian parameter $\gamma$, as 
\be
\gamma \equiv -\frac{\Phi}{\Psi}\,,
\ee
whose experimental bound is 
$|\gamma-1|<2.3 \times 10^{-5}$ \cite{Will}.
{}From Eq.~(\ref{Psiso1}) we have 
$\gamma \simeq 1$ in the present model.
The higher-order terms we neglected to 
derive the solutions (\ref{Psiso1}) are 
even much smaller than the term 
$2Q^2 (r/r_V)^2$.
Hence the experimental bound of $\gamma$ is well 
satisfied inside the Vainshtein radius.

For $r$ close to 0 the solution to Eq.~(\ref{phieq}) is 
different from Eq.~(\ref{phiin2}). 
In this regime there is the solution 
\be
\phi'(r)={\cal C}M^3r\,,
\ee
where ${\cal C}$ is a dimensionless constant determined below. 
We assume that $\rho_m$ approaches a constant value $\rho_c$
as $r \to 0$. Since $\mu_4 \simeq Q/[M_{\rm pl}(1+12c_4{\cal C}^2)]$ 
and $\mu_5 \simeq 24c_4 {\cal C}^3M^3/(1+12c_4{\cal C}^2)$, 
integration of Eq.~(\ref{phieq}) gives the relation 
$3{\cal C} \left( 4c_4 {\cal C}^2+1 \right)
\simeq Q\rho_c/(M^3M_{\rm pl})$. 
If we consider the mass scale $M^3 \approx M_{\rm pl}H_0^2 
\approx \rho_0/M_{\rm pl}$, then 
$|Q\rho_c/(M^3M_{\rm pl})| \approx |Q|\rho_c/\rho_0 \gg 1$ 
for the Sun. 
Since $|4c_4 {\cal C}^2| \gg 1$ in this case, the constant 
${\cal C}$ reduces to 
\be
{\cal C} \simeq \left( \frac{Q\rho_c}{12c_4M^3M_{\rm pl}} 
\right)^{1/3}\,,
\label{Cso}
\ee
by which the solution is given by 
\be
\phi'(r) \simeq \left( \frac{Q\rho_c}{12c_4M_{\rm pl}} \right)^{1/3}
M^2 r
\qquad \quad (r \sim 0)\,.
\label{phiorigin}
\ee
This satisfies the boundary condition (\ref{phicen}).
The sign of (\ref{phiorigin}) should be the same as
(\ref{phiin2}) for the matching of two solutions, 
in which case we require 
\be
c_4>0\,.
\ee

For the star with constant density $\rho_c$ the solution (\ref{phiorigin}) 
should be valid up to the radius $r_s$. 
In this case the Schwarzschild radius is 
$r_g=\rho_c r_s^3/(3M_{\rm pl}^2)$ from Eq.~(\ref{rgdef}). 
Using the Vainshtein radius (\ref{rV1}), the constant ${\cal C}$
in Eq.~(\ref{Cso}) can be estimated as $|{\cal C}| \simeq r_V/r_s$ 
for $c_4={\cal O}(1)$. Then the solution inside the star
is given by $|\phi_{\rm in}'(r)| \simeq (r_V/r_s)M^3r$, 
by which $|\phi_{\rm in}'(r_s)| \simeq M^3r_V$
around the surface. Since the solution outside the star
is $\phi_{\rm out}'(r) \simeq QM_{\rm pl}r_g/r_V^2$, 
we find that $|\phi_{\rm in}'(r_s)|/|\phi_{\rm out}'(r_s)| \simeq 1$ 
by using Eq.~(\ref{rV1}).
Hence the two solutions smoothy connect each other 
around the surface of the body.
In other words, the Vainshtein mechanism is at work for 
the solution (\ref{phiorigin}) inside the body.

In order to see how the matching of the two solutions (\ref{phiin2}) and 
(\ref{phiorigin}) occurs for the varying matter density, 
we solve the field equation (\ref{phieq}) 
numerically for the density profile (\ref{profile}).
We introduce the following dimensionless variables
\be
x=\frac{r}{r_s}\,,\qquad y=\frac{M_{\rm pl}}
{M^6 \rho_c r_s^3}\phi'^3(r)\,,\qquad
z=\frac{\phi}{M_{\rm pl}}\,,\qquad
\beta_t=\frac{r_t}{r_s}\,,\qquad
b_1=\left( \frac{\rho_c r_s^2}{M_{\rm pl}^2} \right)^{1/3}\,,
\qquad
b_2=\left( \frac{M^3 r_s^2}{M_{\rm pl}} \right)^{1/3}\,.
\label{xz}
\ee
The parameter $b_1$ can be estimated as 
$b_1 \approx 0.1$ for the Sun ($\rho_c \approx 100$ g/cm$^3$, 
$r_s \approx 7 \times 10^{10}$ cm) and 
$b_1 \approx 10^{-3}$ for the Earth 
($\rho_c \approx 10$ g/cm$^3$, $r_s \approx 6 \times 10^{8}$ cm), 
respectively. The parameter $b_2$ depends on the mass scale $M$.
If we take the mass $M^3 \approx M_{\rm pl} H_0^2$ relevant to 
dark energy, we have $b_2 \approx 10^{-12}$ for the Sun and 
$b_2 \approx 10^{-13}$ for the Earth.
The field equation (\ref{phieq}) can be rewritten in terms of 
the dimensionless variables (\ref{xz}).
In realistic situations it is a good approximation to neglect the
terms including $b_2$, in which case Eq.~(\ref{phieq}) reads
\be
\frac{dy(x)}{dx} \simeq \frac{1}{4c_4} x 
\left[ Qx+6c_4 b_1^3 e^{2Qz(x)} y(x) \right] e^{-x^2/\beta_t^2}\,.
\label{dyxeqs}
\ee
The variable $z(x)$ obeys the differential equation 
\be
\frac{dz(x)}{dx}=b_1b_2^2\,y(x)^{1/3}\,.
\label{dzx}
\ee

For $x$ close to 0, the solution (\ref{phiorigin}) 
corresponds to $y(x)^{1/3} \simeq [Q/(12c_4)]^{1/3}x$, in which case
the first term on the r.h.s. of Eq.~(\ref{dyxeqs}) dominates 
over the second term.
In the regime $r_g \ll r \ll r_V$ the solution is given 
by Eq.~(\ref{phiin2}), i.e., $y(x)^{1/3}=Qr_sr_g/(b_1b_2^2 r_V^2)$.
For $M^3 \approx M_{\rm pl} H_0^2$ the order of (\ref{phiin2})
be estimated as $y(x)^{1/3}={\cal O}(0.1)$ for both the Sun and 
the Earth. {}From Eq.~(\ref{dzx}) the variation of $z(x)$ is negligibly small,
so that $e^{2Qz(x)} \simeq 1$ in Eq.~(\ref{dyxeqs}).
The second term in the square bracket of Eq.~(\ref{dyxeqs}) becomes
comparable to the term $Qx$ for the distance $x>(2b_1^{-3})^{1/2}$, 
which translates into the condition $r \gg r_s$ for both the Sun and 
the Earth. Outside the star the r.h.s. of Eq.~(\ref{dyxeqs}) starts to
decrease rapidly by the exponential factor $e^{-x^2/\beta_t^2}$.
Then, for $r \gg r_s$, the solution should be described by $y(x)=$\,constant, i.e., 
(\ref{phiin2}).

\begin{figure}
\includegraphics[height=3.3in,width=3.5in]{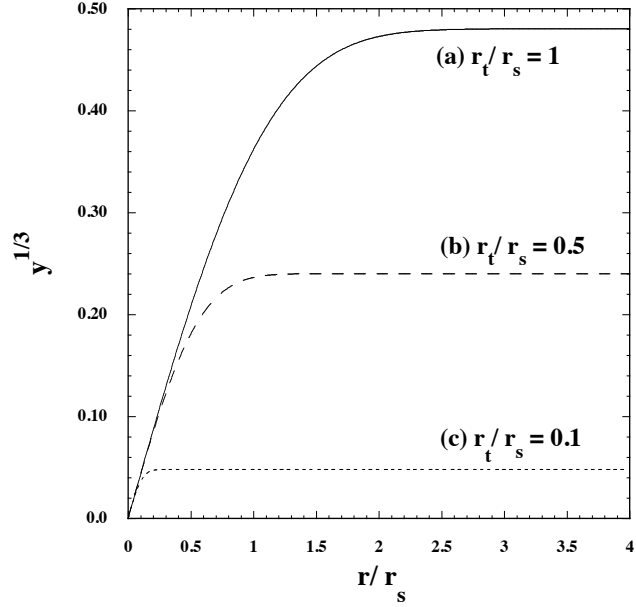}
\caption{\label{fig1}
The field derivative $y^{1/3}=[M_{\rm pl}/(M^6\rho_c r_s^3)]^{1/3} \phi'$ 
versus $r/r_s$ for the matter density profile (\ref{profile}) 
with $c_4=1$, $Q=1$, $b_1=0.1$, 
$b_2=3 \times 10^{-12}$. The boundary conditions
are chosen to be $y(0)=0$ and $z(0)=0$.
Each case corresponds to (a) $r_t/r_s=1$, 
(b) $r_t/r_s=0.5$, and (c) $r_t/r_s=0.1$, respectively.}
\end{figure}

Numerically we solve Eqs.~(\ref{dyxeqs}) and (\ref{dzx}) 
with the boundary conditions $y(0)=0$ and $z(0)=0$.
Although the terms including $b_2$ are neglected in Eq.~(\ref{dyxeqs}), 
we checked that using the full field equation of motion
gives practically identical results
for $b_2 \ll 1$. Figure \ref{fig1} shows $y^{1/3}$ versus $r/r_s$ 
for three different values of $r_t/r_s$. 
Clearly the solution (\ref{phiorigin}) smoothly connects with 
another solution (\ref{phiin2}).
For smaller $r_t/r_s$ the term $\mu_4 \rho_m$ starts to be suppressed
at shorter distances, so that the transition to the solution (\ref{phiin2})
occurs at smaller $r$.
The numerical values of the asymptotically constant solution 
are typically of the order of $y^{1/3}={\cal O}(0.1)$.

Substituting the solution (\ref{phiorigin}) into Eqs.~(\ref{eq:00}) 
and (\ref{eq:11}), we find that the corrections from the field derivative 
to $\Phi$ and $\Psi$ are proportional to $r^2$.
For the star with a nearly constant density 
the leading-order contributions to the gravitational potentials
have the $r$-dependence: $\Phi \propto \Psi \propto r^2$.
The ratio between the corrections and the leading-order terms
is of the order of 
$|{\cal C}QM^3 M_{\rm pl}/\rho_c| \approx Q^2/{\cal C}^2 \approx 
Q^2(r_s/r_V)^2 \ll 1$, so that the corrections are 
suppressed for $r \lesssim r_s \ll r_V$.

\section{Covariant Galileons}
\label{galisec}

The covariant Galileon \cite{Deffayet} is characterized by the Lagrangian
\be
G_3(X)=\frac{c_3}{M^3}X\,,\qquad
G_4 (\phi,X)=\frac{M_{\rm pl}^2}{2} e^{-2Q\phi/M_{\rm pl}}+
\frac{c_4}{M^6}X^2\,,\qquad
G_5(X)=\frac{c_5}{M^9}X^2\,,
\ee
where $c_{3,4,5}$ are dimensionless constants, and 
$M$ is a constant having a dimension of mass. 
Since $G_5(X)$ does not have a $\phi$-dependence, the terms 
such as $G_{5,\phi X}$ and $G_{5,\phi XX}$ in Eqs.~(\ref{mu5})
and (\ref{beta}) vanish. This means that $G_5(X)$ alone 
does not accommodate the Vainshtein mechanism.
This situation is different in the presence of the terms 
$G_3(X)$ and $G_4(\phi,X)$ given above.

{}From Eq.~(\ref{varadius}) the Vainshtein radius $r_V$ is 
given by 
\be
\frac{M^3r_V^3}{QM_{\rm pl}r_g}=c_V\,,
\label{vacv}
\ee
where $c_V=2c_3 \pm \sqrt{4c_3^2-12c_4}$ or 
$c_V=-2c_3 \pm \sqrt{4c_3^2+12c_4}$.
The signs inside $c_V$ should be chosen to have 
a real value of $c_V$ consistent with the l.h.s. of Eq.~(\ref{vacv}).
For $c_3=c_4=1$ and $Q>0$, it follows that 
$r_V=(2Q M_{\rm pl}r_g)^{1/3}/M$.
In the limit that $c_4 \to 0$ and $c_3 \to 0$ we have 
$M^3r_V^3/(QM_{\rm pl}r_g)=\pm 4c_3$ and 
$M^3r_V^3/(QM_{\rm pl}r_g)=\pm 2\sqrt{3|c_4|}$, respectively.
In the following we study the case in which $r_V$ 
is of the order of $(|Q|M_{\rm pl}r_g)^{1/3}/M$, i.e., 
$|c_V| \sim 1$. 

\subsection{$c_5=0$}
\label{c50}

Let us first consider the case in which the term $G_5(X)$ is absent. 
Using the solution (\ref{phila}) in the regime $r \gg r_V$, one can show 
that the conditions (\ref{con1la}), (\ref{con2la}), and (\ref{con4la}) 
are satisfied for $r_V \gg r_g$. 
For the distance $r$ not away from $r_V$ ($r \gtrsim r_V$), 
the quantities $\mu_4$ and $\mu_5$ can be estimated as
\be
\mu_4 \simeq \frac{Q}{M_{\rm pl}}\,,\qquad
\mu_5 \simeq -\frac{6Q^2M_{\rm pl}^2 r_g^2}{M^3 r^6}
\left( c_3 -\frac{4c_4QM_{\rm pl}r_g}{M^3 r^3} \right)\,.
\label{mu45es}
\ee
When $c_3=0$, the distance $r_*$ at which $|\mu_4| \rho_m=|\mu_5|$
is given by Eq.~(\ref{rstar}). 
If $c_4=0$, then we obtain 
$r_*=[6|c_3||Q|M_{\rm pl}^3 r_g^2/(M^3 \rho_m)]^{1/6}
\approx [M^3M_{\rm pl}/(|Q|\rho_m)]^{1/6}r_V$ for $|c_3| \sim 1$.
For $M^3 \approx M_{\rm pl}H_0^2 \approx \rho_0/M_{\rm pl}$
and $|Q|={\cal O}(1)$ it follows that $r_* \approx (\rho_0/\rho_m)^{1/6}r_V$.
If $\rho_m$ is not significantly away from $\rho_0$, $r_*$
is the same order as $r_V$.
Since the second term in the parenthesis of $\mu_5$ in 
Eq.~(\ref{mu45es}) is of the order of $|c_4|(r_V/r)^3$, 
the distance $r_*$ in the case $|c_3| \sim |c_4| \sim 1$ is 
similar to that discussed above for $\rho_m$ not 
significantly different from $\rho_0$.

In the regime $r_g \ll r \ll r_V$ the field equation of motion 
(\ref{fieldap}) reads
\be
\phi''(r)+\frac{\phi'(r)}{2r} \left[ 1-
\frac{3\alpha_{43}\phi'(r)}{M^3r} \right]^{-1} \simeq 0\,,\qquad
{\rm where} \qquad
\alpha_{43} \equiv \frac{c_4}{c_3}\,.
\label{phiL34}
\ee
This is integrated to give
\be
r \phi'^2(r)-\frac{2\alpha_{43}}{M^3}\phi'^3(r)=C\,,
\label{implicit}
\ee
where $C$ is an integration constant determined 
by matching (\ref{implicit}) with the solution 
(\ref{phila}) at $r=r_V$.
Then, the implicit solution (\ref{implicit}) reads
\be
r \phi'^2(r)-\frac{2\alpha_{43}}{M^3}\phi'^3(r)=
\frac{(QM_{\rm pl}r_g)^2}{r_V^3} \left( 
1-\frac{2\alpha_{43}}{c_V} 
\right) \qquad \quad (r \ll r_V).
\label{phiim}
\ee
In the limit $\alpha_{43} \to 0$ we have
$\phi'(r) \propto r^{-1/2}$, whereas for $|\alpha_{43}| \to \infty$
there is the solution $\phi'(r)=$\,constant.
The latter corresponds to the one derived in Eq.~(\ref{phiin}).
The behavior of solutions changes at the radius $r_{43}$ satisfying 
\be
r_{43}=2|\alpha_{43}\phi'(r_{43})|/M^3\,.
\label{r43}
\ee
In the following we study two qualitatively different 
cases separately.

\begin{itemize}
\item (i) $r_{43} \gg r_V$

Let us first consider the case $r_{43} \gg r_V$.
Substituting the solution (\ref{phila}) into 
Eq.~(\ref{r43}), we obtain 
\be
r_{43}=\frac{(2|\alpha_{43}||Q|M_{\rm pl}r_g)^{1/3}}{M}
=\left( \frac{2|\alpha_{43}|}{|c_V|} \right)^{1/3}r_V\,.
\ee
Since $|c_V| \sim 1$
the condition $r_{43} \gg r_V$ translates to $|\alpha_{43}| \gg 1$,
i.e., $|c_3| \ll 1$ and $|c_4| \sim 1$.
For the distance $r \ll r_V$, the dominant contribution 
to the l.h.s. of Eq.~(\ref{phiim}) is the second term. 
The first term of Eq.~(\ref{phiim}) can be treated as 
a perturbation to the leading-order solution. 
In this way we can derive 
the approximate solution
\be
\phi'(r) \simeq \frac{QM_{\rm pl}r_g}{r_V^2}
\left[ 1-\frac{c_V}{6\alpha_{43}} \left(1-\frac{r}{r_V}
\right) \right] \qquad \quad
(r \ll r_V \ll r_{43}).
\label{phidi}
\ee
This case is similar to what we discussed in 
Sec.~\ref{concretemodel}, but there is a correction 
coming from the term $G_3(X)$.
On using $|\phi''(r)| \approx |c_V \phi'(r)/(6\alpha_{43}r_V)|$,
one can show that this correction is negligibly small 
in Eqs.~(\ref{eq:00}) and (\ref{eq:11}).
Then the gravitational potentials are approximately 
given by Eq.~(\ref{Psiso1}), so that 
local gravity constraints are satisfied deep 
inside the Vainshtein radius.

\item (ii) $r_{43} \ll r_V$

In another case $r_{43} \ll r_V$, the first term on the l.h.s. of 
Eq.~(\ref{phiim}) is the dominant contribution for the distance  
$r_{43} \ll r \ll r_V$. Dealing with the second term of Eq.~(\ref{phiim}) 
as a perturbation to the leading-order solution of $\phi'(r)$, 
it follows that 
\be
\phi'(r) \simeq \frac{QM_{\rm pl}r_g}{r_V^{3/2}r^{1/2}}
\left[ 1-\frac{\alpha_{43}}{c_V}
\left\{ 1-\left( \frac{r_V}{r} \right)^{3/2} \right\} \right]
\qquad \quad
(r_{43} \ll r \ll r_V).
\label{phiel}
\ee
The leading-order solution $\phi'(r)=QM_{\rm pl}r_g/(r_V^{3/2}r^{1/2})$
is the same as that derived in Ref.~\cite{VainARS} in the presence of the 
term $G_3$ alone.
{}From Eq.~(\ref{r43}) the distance $r_{43}$ can be estimated as 
\be
r_{43} \simeq \frac{(2|\alpha_{43}||Q|M_{\rm pl}r_g)^{2/3}}{M^2r_V}
=\left( \frac{2|\alpha_{43}|}{|c_V|} \right)^{2/3}r_V\,.
\label{r43es}
\ee
The condition $r_{43} \ll r_V$ translates to $|\alpha_{43}| \ll 1$, 
i.e., $|c_4| \ll 1$ and $|c_3| \sim 1$.
In the regime $r \ll r_{43}$ the dominant contribution to the 
l.h.s. of Eq.~(\ref{phiim}) is the second term. Taking into 
account the first term of Eq.~(\ref{phiim}) 
as a perturbation, we obtain the following solution 
\be
\phi'(r) \simeq \frac{M(QM_{\rm pl}r_g)^{2/3}}
{(-2\alpha_{43})^{1/3} r_V} \left[ 1-\frac{2\alpha_{43}}{3c_V}
-\frac13 \left( \frac{c_V}{-2\alpha_{43}} \right)^{2/3} 
\frac{r}{r_V} \right]
\qquad \quad
(r \ll r_{43}).
\label{phies}
\ee
{}From Eq.~(\ref{r43es}) the last term in the parenthesis of 
Eq.~(\ref{phies}) is of the order of $r/(3r_{43})$, so that 
it is suppressed in the regime $r \ll r_{43}$. 
The sign of $\phi'(r)$ should not change around $r=r_{43}$, so that 
we require the following condition 
\be
\alpha_{43} Q<0\,.
\label{alcon}
\ee
One can confirm that Eqs.~(\ref{phiel}) and (\ref{phies}) satisfy 
the conditions (\ref{con1sm}), (\ref{con2sm}), and (\ref{con3sm}).

On using Eq.~(\ref{phiel}) in the regime $r_{43} \ll r \ll r_V$, 
Eqs.~(\ref{eq:00}) and (\ref{eq:11}) are approximately given by 
\ba
\frac{d}{dr} (r \Phi) &\simeq& -\frac{3Q\phi' r}{2M_{\rm pl}}
+\frac{\rho_m r^2}{2M_{\rm pl}^2}\,,\label{G3Phi} \\
\Psi' &\simeq& \frac{\Phi}{r}+\frac{2Q\phi'}{M_{\rm pl}}\,.
\label{G3Psi}
\ea
Substituting the approximate solution 
$\phi'(r) \simeq QM_{\rm pl}r_g/(r_V^{3/2}r^{1/2})$ into 
Eqs.~(\ref{G3Phi}) and (\ref{G3Psi}), 
the integrated solutions are 
\be
\Phi \simeq \frac{r_g}{2r} \left[ 1-2Q^2 \left(
\frac{r}{r_V} \right)^{3/2} \right]\,,\qquad
\Psi \simeq -\frac{r_g}{2r} \left[ 1-4Q^2 \left(
\frac{r}{r_V} \right)^{3/2} \right]
\qquad (r_{43} \ll r \ll r_V). \label{Psiso2}
\ee
For the distance $r$ close to $r_{43}$ the correction term from 
$\alpha_{43}$ in Eq.~(\ref{phiel}) tends to be important, but this does not 
change the order of the estimation (\ref{Psiso2}).
Since $2Q^2 (r/r_V)^{3/2} \ll 1$ deep inside the Vainshtein radius, 
the deviation of the post-Newtonian parameter 
$\gamma=-\Phi/\Psi$ from 1 is much smaller than unity.

Employing the solution (\ref{phies}) in the regime $r \ll r_{43}$,
the gravitational potentials approximately satisfy Eqs.~(\ref{Phieq}) 
and (\ref{Psieq}) with $\phi'(r) \simeq M(QM_{\rm pl}r_g)^{2/3}
/[(-2\alpha_{43})^{1/3}r_V] \simeq QM_{\rm pl}r_g/(r_V^{3/2} r_{43}^{1/2})$
for $r \gg r_g$, where we used the condition (\ref{alcon}). 
Then it follows that 
\be
\Phi \simeq \frac{r_g}{2r} \left( 1-2Q^2
\frac{r^2}{r_V^{3/2} r_{43}^{1/2}}
\right)\,,\qquad
\Psi \simeq -\frac{r_g}{2r} \left( 1-2Q^2
\frac{r^2}{r_V^{3/2} r_{43}^{1/2}}
\right)
\qquad (r \ll r_{43})\,, \label{Psiso3}
\ee
from which $\gamma \simeq 1$.
The correction terms in Eq.~(\ref{phies}) only give the contributions 
much smaller than the term $2Q^2 r^2/(r_V^{3/2} r_{43}^{1/2})$ ($\ll 1$), 
so that the experimental bound of $\gamma$ is well satisfied.

\end{itemize}

If $|c_3| \sim |c_4| \sim 1$, then $r_{43}$ is the same order as 
$r_V$. In this case the regime in which the $G_3$ term
contributes to the field equation for $r<r_V$
is narrow, so that the solution is described by the 
$G_4$-dominant one ($\phi'(r)={\rm constant}$) for most of $r$ 
smaller than $r_V$. The gravitational potentials within 
the Vainshtein radius can be estimated by taking the limit 
$r_{43} \to r_V$ in Eq.~(\ref{Psiso3}), i.e., Eq.~(\ref{Psiso1}).

As we studied in Sec.~\ref{concretemodel}, both 
(\ref{phidi}) and (\ref{phies}) can connect with another solution 
(\ref{phiorigin}) around the surface of the star.
There is an extreme case $r_{43} \to 0$,
in which the field derivative is given by Eq.~(\ref{phiel}) 
even for small $r$ down to the radius of the star.
In such a case we discuss the matching of solutions 
for more general models in Sec.~\ref{exgali}.

\subsection{$c_5\neq 0$}
\label{c5neq}

We estimate the effect of the term $G_5(X)=c_5X^2/M^9$ 
on the solutions discussed in Sec.~\ref{c50}.
We study two qualitatively different cases: 
(1) $|c_4| \sim 1$, $|c_3| \ll 1$, and 
(2) $|c_3| \sim 1$, $|c_4| \ll 1$.

\subsubsection{$|c_4| \sim 1$, $|c_3| \ll 1$}

This corresponds to the case (i) studied in Sec.~\ref{c50}. 
In the following we focus on the case in which the effect 
of the $c_3$ term is practically absent, i.e., the 
limit $|\alpha_{43}| \to \infty$. 
At small $r$ the term $|4X(G_{5,X}+XG_{5,XX})/r^2|$ 
in Eq.~(\ref{mu4}) gets larger than the other term 
$|2G_{4,\phi}|$. Employing the solution $\phi'(r) 
\simeq QM_{\rm pl}r_g/r_V^2$, this region is estimated as 
$r<(2|c_5|r_gr_V/|c_V|^3)^{1/2}\simeq(r_gr_V)^{1/2}$ 
for $|c_5| \sim 1$.
For the mass scale $M^3 \sim M_{\rm pl}H_0^2$, 
this condition translates to $r\lesssim10^{13}$~cm 
for the Sun and $r\lesssim10^{9}$~cm for the Earth.
In the regime $r\lesssim (r_gr_V)^{1/2}$ we have that 
$\mu_4 \simeq c_5Q^2r_g^2/(6c_4M^3r_V^4)$ 
and $\mu_5\simeq 2QM_{\rm pl}r_g/(r_V^2r)$. 
Since the density $\rho_m(r)$ grows around the surface 
of the star, the condition $|\mu_4|\rho_m(r)<|\mu_5|$ 
can be violated for 
\be
r\rho_m(r) >12\left|\frac{c_4}{c_5 Q}\right|\frac{M^3 M_{\rm pl} r_V^2}{r_g} 
\approx \frac{M_{\rm pl}^2}{r_V}\,,
\ee
where the second approximate equality holds for 
$|c_5| \sim 1$.
Around the surfaces ($r=r_s$) of the Sun and the Earth the 
term $r_s \rho_m (r_s)$ becomes the same order 
as $M_{\rm pl}^2/r_V$ for $\rho_m(r_s) \approx 10^{-5}$ g/cm$^3$
and $\rho_m(r_s) \approx 0.1$ g/cm$^3$, respectively.
Inside these stars we have $|\mu_4|\rho_m(r)>|\mu_5|$, 
so that the solution $\phi'(r) 
\simeq QM_{\rm pl}r_g/r_V^2$ is subject to change.

Outside the star ($r>r_s$), let us estimate the effect of 
the $c_5$ term on the gravitational potentials.
Substituting the solution $\phi'(r) \simeq QM_{\rm pl}r_g/r_V^2$
into Eqs.~(\ref{eq:00}) and (\ref{eq:11}), we obtain
\be
\Phi \simeq \frac{r_g}{2r} \left[ 1-2Q^2 \left(
\frac{r}{r_V} \right)^2-\frac{3c_5 Q^2 r_g^2}{c_V^3r_V r} 
\right]\,, \qquad
\Psi \simeq -\frac{r_g}{2r} \left[ 1-2Q^2 \left(
\frac{r}{r_V} \right)^2-\frac{3c_5 Q^2 r_g^2}{c_V^3r_V r} 
\right]\,,
\ee
where we used the fact that the dominant contribution to $\Phi'$
is $-r_g/(2r^2)$. The corrections to $\Phi$ and $\Psi$ 
coming from the $c_5$ term are negligibly small for
$r>r_s$ (at most of the order of $10^{-21}$ for the Sun).

\begin{figure}
\includegraphics[height=3.1in,width=3.3in]{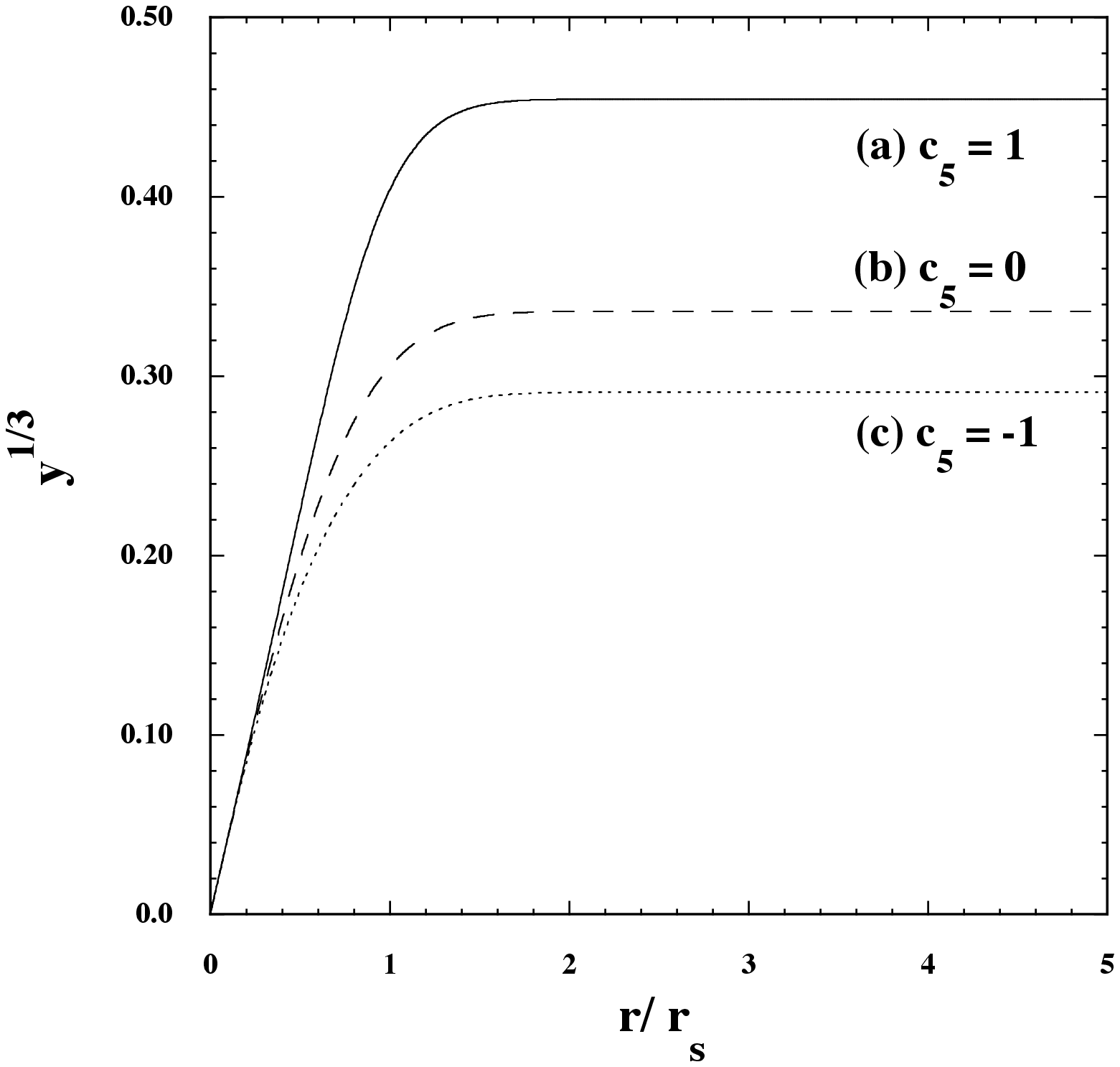}
\includegraphics[height=3.1in,width=3.3in]{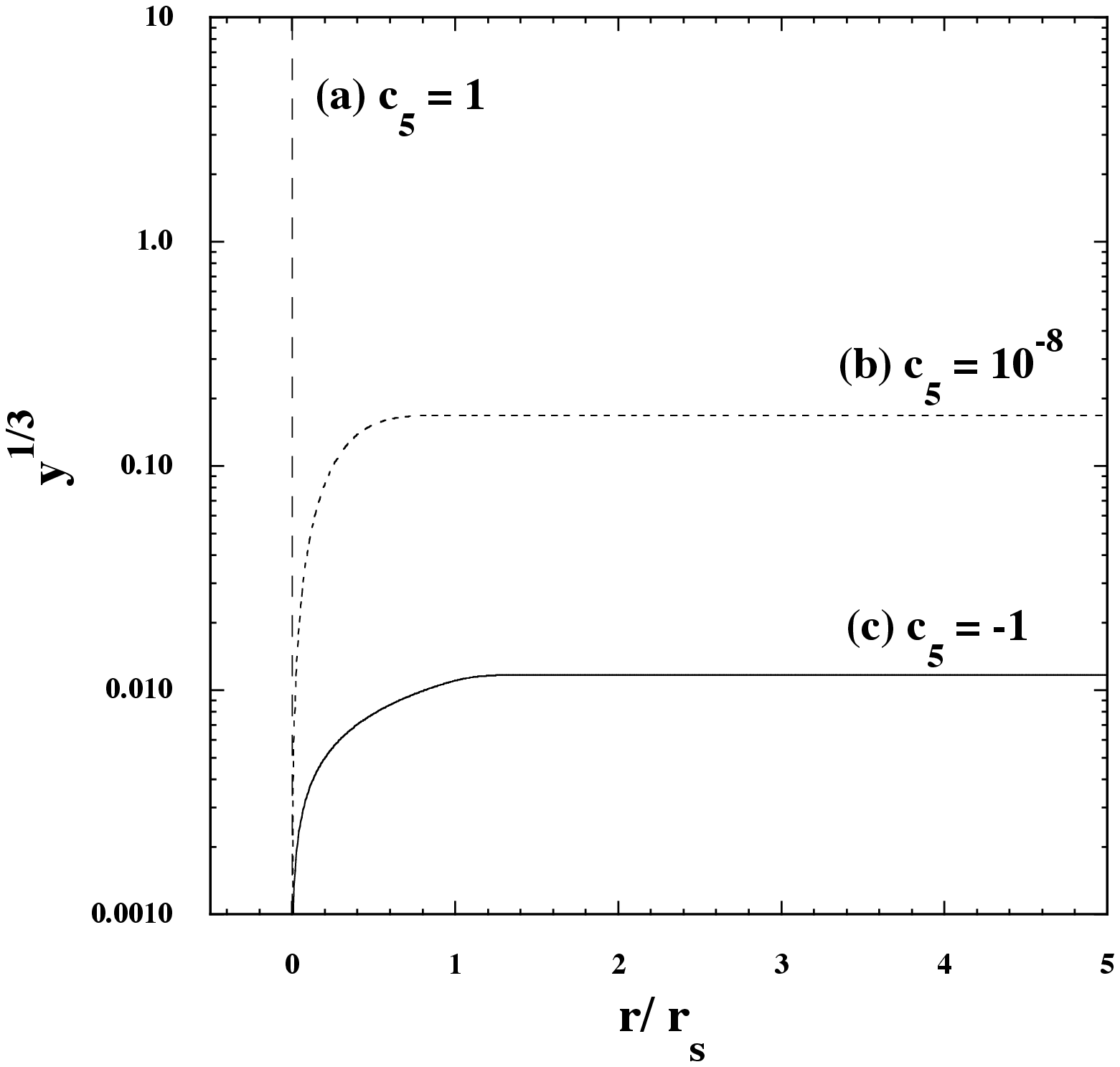}
\caption{\label{fig2}
The field derivative $y^{1/3}=[M_{\rm pl}/(M^6 \rho_c r_s^3)]^{1/3} \phi'$ 
versus $r/r_s$ for the density profile (\ref{profile}). 
The model parameters are $c_4=1$ and $Q=1$ 
with the boundary conditions $y(0)=0$ and $z(0)=0$. 
The left panel corresponds to 
$b_1=1.6 \times 10^{-3}$, $b_2=1.6 \times 10^{-13}$, 
$\beta_t=0.7$ with three different values of $c_5$, whereas 
in the right panel the model parameters are
$b_1=0.1$, $b_2=3.0 \times 10^{-12}$, 
$\beta_t=0.35$ with three different values of $c_5$.}
\end{figure}

Inside the star we study the solution of Eq.~(\ref{phieq}) to see 
whether the matching with another solution $\phi'(r) 
\simeq QM_{\rm pl}r_g/r_V^2$ can be done properly. 
We use the density profile (\ref{profile}) together with 
the dimensionless variables defined in Eq.~(\ref{xz}).
Under the approximation that the terms including $b_2$ 
are negligible, the field equation (\ref{phieq}) reads
\be
\frac{dy(x)}{dx} \simeq \frac{1}{4c_4} 
\left[ Qx^2+6c_4 b_1^3 e^{2Qz(x)} x y(x)+
2c_5 (b_1^4/b_2) e^{2Qz(x)}  y(x)^{4/3} \right] e^{-x^2/\beta_t^2}\,,
\label{dyxeqs2}
\ee
where the variable $z(x)$ satisfies the same equation 
as (\ref{dzx}). 
In the following we study the case $c_4>0$ and $Q>0$.
Around $x \sim 0$ the solution of 
Eq.~(\ref{dyxeqs2}) 
is given by $y(x)^{1/3} \simeq [Q/(12c_4)]^{1/3}x$. 
The third term on the r.h.s. of Eq.~(\ref{dyxeqs2}) dominates over the 
first one for $x^2>x_*^2 \equiv (12c_4)^{4/3}b_2/(2|c_5| Q^{1/3}b_1^4)$.
As long as $x_*$ is larger than 1, the effect of the $c_5$ term
does not manifest itself inside the star. 
This demands the following condition 
\be
|c_5|< \frac{(12c_4)^{4/3}}{2Q^{1/3}} \frac{b_2}{b_1^4}
\approx \frac{10b_2}{b_1^4}=
\frac{10MM_{\rm pl}^{7/3}}{r_s^2 \rho_c^{4/3}}\,.
\label{c5condi}
\ee
If $M^3 \approx M_{\rm pl}H_0^2$, then we have $|c_5| \lesssim 1$
for the Earth ($b_1 \approx 10^{-3}$ and $b_2 \approx 10^{-13}$)
and $|c_5| \lesssim 10^{-7}$ for the Sun ($b_1 \approx 0.1$ and 
$b_2 \approx 10^{-12}$).
Thus the upper bound of $|c_5|$ depends on the density and 
the radius of the star.

Numerically we solve the field equation of motion (\ref{phieq}) 
without neglecting the terms including $b_2$.
In the left panel of Fig.~\ref{fig2} we plot the field derivative $y^{1/3}$ 
versus $r/r_s$ for $b_1=1.6 \times 10^{-3}$, $b_2=1.6 \times 10^{-13}$, 
and $\beta_t=0.7$ with three different values of $c_5$.
This case mimics the density profile of the Earth.
Even for the cases $c_5=1$ and $c_5=-1$, the solution inside 
the star smoothly connects to the exterior 
solution $y^{1/3}={\cal O}(0.1)$.

The right panel of Fig.~\ref{fig2} corresponds to the model 
parameters $b_1=0.1$, $b_2=3.0 \times 10^{-12}$, 
and $\beta_t=0.35$, in which case the density profile is 
similar to that of the Sun.
For the values of $c_5$ satisfying the condition (\ref{c5condi}), e.g., 
the case (b) in Fig.~\ref{fig2}, the matching of the interior and 
exterior solutions occurs smoothly.
If $c_5 \sim 1$, however, the third term on the r.h.s. of 
Eq.~(\ref{dyxeqs2}) dominates over the first term for $r \ll r_s$.
This leads to the rapid growth of $\phi'(r)$ at small $r$, 
in which case the matching with another solution 
$y^{1/3}={\cal O}(0.1)$ outside the star does not occur.
In Fig.~\ref{fig2} this behavior is clearly seen in the 
case $c_5=1$. 

For negative values of $c_5$ satisfying the condition 
$|c_5| \gg 10b_2/b_1^4$, we numerically confirmed that 
the first term on the r.h.s. of Eq.~(\ref{dyxeqs2})
almost balances with the third term, i.e., 
$y(x)^{1/3} \simeq [Qb_2/(2|c_5|b_1^4)]^{1/4}x^{1/2} \ll x^{1/2}$
for $r \lesssim r_s$.
For larger $|c_5|$ the field derivative $y(x)^{1/3}$ gets smaller
around the surface of the star. 
As we see in the case (c) of Fig.~\ref{fig2}, 
we have $y(1)^{1/3}={\cal O}(10^{-2})$ for $c_5 \sim -1$.
This is by one order of magnitude smaller than the exterior solution 
$y^{1/3}={\cal O}(0.1)$ and hence there is a 
problem of the matching of two solutions.

In summary, for the values of $c_5$ satisfying the condition (\ref{c5condi}), 
the solution inside the star smoothly connects to the exterior
solution $\phi'(r) \simeq QM_{\rm pl}r_g/r_V^2$ around 
the surface. The upper bound of $|c_5|$ depends on 
the radius and density of the star.

\subsubsection{$|c_3| \sim 1$, $|c_4| \ll 1$}

This belongs to the case (ii) discussed in Sec.~\ref{c50}.
We study the case in which the effect of the $c_4$ term is 
practically absent, i.e., $\alpha_{43} \to 0$, so that the 
solution in the regime $r_g \ll r \ll r_V$ is given by 
$\phi'(r) \simeq QM_{\rm pl}r_g/(r_V^{3/2} r^{1/2})$.
The term $|4X(G_{5,X}+XG_{5,XX})/r^2|=4|c_5|\phi'^4/(M^9r^2)$ 
in Eq.~(\ref{mu4}) becomes larger than the other term $|2G_{4,\phi}|$
for the distance $r<(2|c_5|r_gr_V^3/|c_V^3|)^{1/4}$.
For the Sun with $|c_5| \sim 1$ and $M^3 \approx M_{\rm pl}H_0^2$, 
this condition translates to $r \lesssim 10^{17}$~cm.
Using the solution $\phi'(r) \simeq QM_{\rm pl}r_g/(r_V^{3/2} r^{1/2})$ 
in the regime $r \lesssim (r_g r_V^3)^{1/4}$, 
we have that 
$\mu_4 \simeq -c_5Qr_g r_V^{3/2}/(2c_3c_V^2M_{\rm pl}r^{5/2})$ 
and $\mu_5 \simeq 3QM_{\rm pl}r_g/(r_V^{3/2}r^{3/2})$. 
Then, the condition $|\mu_4|\rho_m<|\mu_5|$ is satisfied 
for the distance
\be
r>r_5 \equiv \frac{|c_5|}{3|c_3|c_V^2} \frac{\rho_m r_V^3}
{M_{\rm pl}^2} \approx  \frac{\rho_m r_V^3}
{M_{\rm pl}^2} \approx 
\frac{\rho_m}{\rho_0} \left( \frac{r_V}{H_0^{-1}}
\right)^2 r_V\,,
\label{rmu5}
\ee
where the second and third approximate equalities are valid for $|c_5| \sim 1$.
If the Vainshtein radius is $r_V \approx 10^{20}$~cm, 
it follows that $r_5 \approx 10^4 (\rho_m/\rho_0)$~cm for $|c_5| \sim 1$.
The lower bound of $r$ depends on the density profile of 
the star. If we use the mean density $\rho_m \approx 10^{-24}$~g/cm$^3$
of our galaxy, the condition (\ref{rmu5}) corresponds to 
$r>10^9$~cm (whose lower bound is of the same order 
as the radius of the Earth).
Around the Sun the density $\rho_m$ is much larger
than $10^{-24}$~g/cm$^3$, so that the condition (\ref{rmu5})
translates to $r \gg 10^9$~cm.
This suggests that the condition $|\mu_4|\rho_m<|\mu_5|$
can be violated in the solar system.

In order to understand how the effect of the $G_5(X)$ term
manifests itself in the regime $|\mu_4|\rho_m<|\mu_5|$, i.e.,
for the radius $r_5 \lesssim r \ll r_V$,
we estimate the behavior of the gravitational potentials 
by employing the solution $\phi'(r) \simeq QM_{\rm pl}r_g/(r_V^{3/2} r^{1/2})$.
Since the leading-order gravitational potentials are 
given by $\Phi \simeq r_g/(2r)$ and $\Psi \simeq -r_g/(2r)$,
the $G_5(X)$-dependent term inside $A_6$ of Eq.~(\ref{eq:00}) 
provides a much larger contribution 
relative to the term $A_3\Phi'/r^2$ for $r \gg r_g$.
Then the r.h.s. of Eq.~(\ref{G3Phi}) gets corrected by the term
$-c_5 Q^2 r_g^2 r_V^{3/2}/(c_V^3 r^{7/2})$, whereas
Eq.~(\ref{G3Psi}) is unchanged.
Integration of these equations gives
\be
\Phi \simeq \frac{r_g}{2r} \left[ 1-2Q^2
\left( \frac{r}{r_V} \right)^{3/2}
+\frac{4c_5Q^2}{5c_V^3} \frac{r_gr_V^{3/2}}{r^{5/2}}
\right]\,,\qquad
\Psi \simeq 
-\frac{r_g}{2r} \left[ 1-4Q^2
\left( \frac{r}{r_V} \right)^{3/2}
+\frac{8c_5Q^2}{35c_V^3} \frac{r_gr_V^{3/2}}{r^{5/2}}
\right]\,.\label{5core2}
\ee

The third terms in Eq.~(\ref{5core2}) dominate over 
the leading-order contribution for the distance 
\be
r<\left( \frac{|c_5|Q^2 r_gr_V^{3/2}}
{|c_V|^3} \right)^{2/5}
\approx (c_5^2 r_g^2 r_V^3)^{1/5}\,.
\label{r5}
\ee
For the Sun with $|c_5| \sim 1$ and $M^3 \approx M_{\rm pl}H_0^2$, 
the condition (\ref{r5}) corresponds to $r<10^{14}$~cm.
Then the experimental bound of the post-Newtonian parameter 
$\gamma$ is not satisfied in the solar system.
Hence the presence of the term $G_5(X)=c_5X^2/M^9$ 
disrupts the Vainshtein mechanism induced by 
the field self-interaction $G_3(X)=c_3X/M^3$. 
For the consistency with local gravity constraints we
require that $|c_5|$ is very much smaller than 1.

\section{Application to other models}
\label{appsec}

In this section we study how the Vainshtein mechanism is at work 
for several concrete models such as (A) extended Galileons, 
(B) Galileons with dilatonic couplings, and 
(C) DBI Galileons with Gauss-Bonnet and other terms.

\subsection{Extended Galileons}
\label{exgali}

The extended Galileon model \cite{KY,exGAL} is given by the Lagrangian 
\be
G_3(X)=c_3 M^{1-4p_3} X^{p_3}\,,\qquad
G_4 (\phi,X)=\frac{M_{\rm pl}^2}{2} e^{-2Q\phi/M_{\rm pl}}+
c_4M^{2-4p_4}X^{p_4}\,,
\ee
where $p_3$ and $p_4$ are integers satisfying
$p_3 \geq 1$ and $p_4 \geq 2$. 
We do not take into account the term 
$G_5(X)=c_5 M^{-1-4p_5}X^{p_5}$ ($p_5 \ge 2$) 
because its effect is similar to what we studied 
in Sec.~\ref{c5neq}.

From Eq.~(\ref{varadius}) the Vainshtein radius can be estimated as
\ba
r_V &\simeq& (|Q|M_{\rm pl} r_g)^{\frac{2p_3-1}{4p_3-1}}/M 
\qquad (|c_3| \sim 1, c_4=0)\,,\label{rVex1} \\
r_V &\simeq& (|Q|M_{\rm pl} r_g)^{\frac{p_4-1}{2p_4-1}}/M 
\qquad (|c_4| \sim 1, c_3=0)\,.\label{rVex2}
\ea
If $p_4=2p_3$, then both (\ref{rVex1}) and (\ref{rVex2}) are the same.
For $p_3 \gg 1$ and $p_4 \gg 1$ it follows that 
$r_V \simeq (|Q|M_{\rm pl} r_g)^{1/2}/M$. 
In the regime $r_g \ll r \ll r_V$, integration of 
Eq.~(\ref{fieldap}) leads to the following implicit solution
\ba
& & p_3 r\phi'(r)^{2p_3}-(-2)^{p_3-p_4+1}p_4 (p_4-1) \alpha_{43}
M^{1+4p_3-4p_4}\phi'(r)^{2p_4-1} \nonumber \\
& &=(QM_{\rm pl}r_g)^{2p_3} r_V^{1-4p_3}
\left[ p_3-(-2)^{p_3-p_4+1}p_4 (p_4-1) \alpha_{43}
(Mr_V)^{1+4p_3-4p_4} (QM_{\rm pl}r_g)^{-1-2p_3+2p_4}
\right]\,,
\label{imgene}
\ea
where $\alpha_{43}=c_4/c_3$ and we used 
Eq.~(\ref{phila}) to match the solutions at $r=r_V$.
For the large distance (or the limit $|\alpha_{43}| \to 0$) 
the solution behaves as $\phi'(r) \propto r^{-1/(2p_3)}$, 
whereas for small $r$ (or the limit $|\alpha_{43}| \to \infty$)
we have $\phi'(r)=$\,constant. 
The behavior of $\phi'(r)$ changes at 
the distance $r_{43}$ satisfying 
\be
r_{43}=p_3^{-1} p_4 (p_4-1)
M^{1+4p_3-4p_4}
|(-2)^{p_3-p_4+1} \alpha_{43} 
\phi'(r_{43})^{2p_4-2p_3-1}|\,.
\label{r43d}
\ee

If $r_{43} \gg r_V$, i.e. $|\alpha_{43}| \gg 1$, then 
the solution in the regime $r_g\ll r \ll r_V$ is
\be
\phi'(r) \simeq \frac{QM_{\rm pl}r_g}{r_V^2}\,,
\label{phid0}
\ee
where we neglected the correction from the $1/\alpha_{43}$ term.
In this case the gravitational potentials are given 
by Eq.~(\ref{Psiso1}) and hence local gravity constraints 
are well satisfied for the distance much smaller than $r_V$.

If $r_{43} \ll r_V$, i.e. $|\alpha_{43}| \ll 1$, we can neglect 
the term including $\alpha_{43}$ on the r.h.s. of Eq.~(\ref{imgene}).
Then we obtain 
\ba
\phi'(r) &\simeq& QM_{\rm pl}r_g 
r_V^{\frac{1-4p_3}{2p_3}} r^{-\frac{1}{2p_3}} \qquad 
(r_{43} \ll r  \ll r_V)\,,\label{phid1}\\
\phi'(r) &\simeq& QM_{\rm pl}r_g r_V^{\frac{1-4p_3}{2p_3}} 
r_{43}^{-\frac{1}{2p_3}} \qquad
(r \ll r_{43})\,.
\label{phid2}
\ea
On using these solutions, Eqs.~(\ref{eq:00}) and (\ref{eq:11}) 
are integrated to give
\ba
& & \Phi \simeq 
\frac{r_g}{2r} \left[ 1-2Q^2 \left( \frac{r}{r_V} \right)
^{\frac{4p_3-1}{2p_3}} \right]\,,
\qquad 
\Psi \simeq 
-\frac{r_g}{2r} \left[ 1-\frac{4p_3}{2p_3-1}
Q^2 \left( \frac{r}{r_V} \right)
^{\frac{4p_3-1}{2p_3}} \right]
\qquad 
(r_{43} \ll r \ll r_V),\label{Phiex1} \\
& & \Phi \simeq 
\frac{r_g}{2r} \Biggl( 1-2Q^2 
\frac{r^2}{r_V^{\frac{4p_3-1}{2p_3}} 
r_{43}^{\frac{1}{2p_3}}}  \Biggr)\,,
\qquad 
\Psi \simeq 
-\frac{r_g}{2r} \Biggl( 1-2Q^2 
\frac{r^2}{r_V^{\frac{4p_3-1}{2p_3}} 
r_{43}^{\frac{1}{2p_3}}}  \Biggr)
\qquad 
(r \ll r_{43}). \label{Psiex1}
\ea
If $p_3 \gg 1$, then both (\ref{Phiex1}) and (\ref{Psiex1}) reduce 
to the result (\ref{Psiso1}). 
This shows that, for larger $p_3$, the deviation from GR 
tends to be smaller.

\begin{figure}
\includegraphics[height=3.3in,width=3.5in]{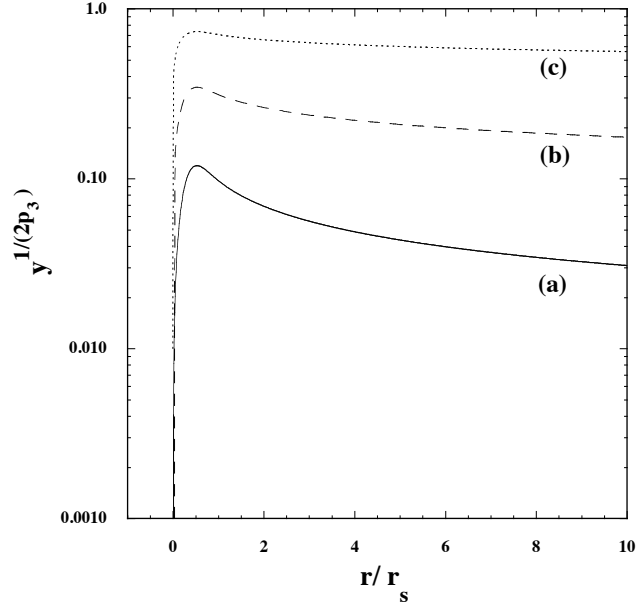}
\caption{\label{fig3}
The field derivative $y^{1/(2p_3)}=
[M_{\rm pl}/(M^{4p_3-1}\rho_c r_s^2)]^{1/(2p_3)} 
\phi'$ versus $r/r_s$ for the density profile (\ref{profile}).
The model parameters are 
$Q=1$, $b_1=0.1$, and 
$r_t/r_s=0.35$ with the boundary conditions $y(0)=0$ and $z(0)=0$.
Each case corresponds to (a) $p_3=1$, $c_3=-1$, 
(b) $p_3=2$, $c_3=1$, and (c) $p_3=5$, $c_3=-1$, 
respectively.}
\end{figure}

Inside the star the above solutions are subject to change.
Let us consider the limit $r_{43} \to 0$, i.e., the case in which 
the solution is given by Eq.~(\ref{phid1}) for small $r$ down to
the surface of the star.
We consider the density profile (\ref{profile}) of the star and 
introduce the following dimensionless variables
\be
y=\frac{M_{\rm pl}}{M^{4p_3-1} \rho_c r_s^2} \phi'^{2p_3}(r)\,,
\qquad b_2=\left( \frac{M^{4p_3-1} r_s^{2p_3}}{M_{\rm pl}^{2p_3-1}}
\right)^{1/4}\,,
\ee
where $x$, $z$, $\beta_t$, and $b_1$ are the same as those 
defined in Eq.~(\ref{xz}).
For the mass scale $M$ relevant to dark energy, we have 
$b_2 \ll 1$ for both the Sun and the Earth.
Neglecting the contribution of the terms including $b_2$, 
the field equation of motion (\ref{phieq}) reads
\be
\frac{dy}{dx} \simeq \frac{1}{4c_3p_3x} \left[ \left\{
(-2)^{p_3}Q+c_3p_3(1+4p_3)b_1^3 y \right\}
x^2 e^{-x^2/\beta_t^2}-4c_3p_3 y \right]\,,
\label{dyx2}
\ee
where we used the approximation $e^{2Qz(x)} \simeq 1$. 
The variable $z(x)$ satisfies the differential equation 
$dz(x)/dx=(b_1^3 b_2^4 y(x))^{1/(2 p_3)}$. 
For small $x$ the second term on the r.h.s. of 
Eq.~(\ref{dyx2}) can be neglected relative to 
other two terms.
Using the approximation $e^{-x^2/\beta_t^2} \simeq 1$ 
in this regime, we obtain the following solution 
\be
y(x) \simeq \frac{(-2)^{p_3}Q}{12c_3p_3}x^2\,,
\label{yxin}
\ee
which means that the field derivative grows as 
$\phi'(r) \propto r^{1/p_3}$. The second term 
on the r.h.s. of Eq.~(\ref{dyx2}) becomes important 
only for $x^2>12/[(1+4p_3)b_1^3]$.
Outside the star the last term in Eq.~(\ref{dyx2}) 
is the dominant contribution, in 
which case the solution is given by $y(x) \propto x^{-1}$. 
In fact this corresponds to Eq.~(\ref{phid1}), i.e.,
$\phi'(r) \propto r^{-1/(2p_3)}$.
In order to match this solution with (\ref{yxin}), 
we require that $c_3Q<0$ for odd $p_3$ and
$c_3Q>0$ for even $p_3$. 

In Fig.~\ref{fig3} we plot the numerically integrated solutions 
of the field derivative for three different values of $p_3$. 
This shows that the solution $\phi'(r) \propto r^{1/p_3}$ 
connects with another one $\phi'(r) \propto r^{-1/(2p_3)}$  
around the surface of the star. 
For $p_3 \gg 1$ the field derivative outside the body is given by 
$\phi'(r) \simeq $\,\,constant, in which case the solution (\ref{phid1}) 
reduces to $\phi'(r) \simeq QM_{\rm pl}r_g/r_V^2$. 

For the star with a nearly constant density the field derivative 
inside the body is estimated as 
$\phi'(r) \simeq (QM_{\rm pl}r_g/r_V^2)
(r_s/r_V)^{-1/(2p_3)}(r/r_s)^{1/p_3}$ by matching the 
two solutions at $r=r_s$.
Substituting this solution into Eqs.~(\ref{eq:00}) and (\ref{eq:11}), 
we find that the corrections from the field derivative to 
$\Phi$ and $\Psi$ are
suppressed under the condition $|Q\phi'| \ll \rho_m r/M_{\rm pl}$.
When $p_3=1$ this condition translates to 
$Q^2(r_s/r_V)^{3/2} \ll 1$, which is well satisfied 
for $r_s \ll r_V$. If $p_3 \geq 2$ and $|Q|={\cal O}(1)$, 
the corresponding condition is given by 
$r \gg \tilde{r} \equiv r_s (r_s/r_V)^{(4p_3-1)/(2p_3-2)}$.
Provided that $r_s \ll r_V$, $\tilde{r}$ is much smaller than $r_s$, 
e.g., $\tilde{r} \simeq 10^{-21}$~cm for $p_3=2$, 
$r_s=7 \times 10^{10}$~cm and $r_V=10^{20}$~cm.
Then, the corrections to the leading-order gravitational 
potentials are suppressed for most of the region inside the star.

In the case where the field derivative is given by either (\ref{phid0}) 
or (\ref{phid2}) outside the star, we confirmed that it smoothly 
connects with another solution inside the star for an appropriate choice
of the sign of $c_4$ (i.e., $c_4<0$ for odd $p_4$ and 
$c_4>0$ for even $p_4$).
Hence the extended Galileon model with $p_3 \geq 1$ and 
$p_4 \geq 2$ can successfully accommodate the Vainshtein 
mechanism with a proper matching of solutions around 
the surface of the star.

\subsection{Galileons with dilatonic couplings}
\label{conformalsec}

Next we proceed to the model in which the $G_{3,4,5}$ terms have 
the dilatonic coupling of the form
\be
G_3(\phi,X)=\frac{c_3}{M^3}e^{-\lambda_3 \phi/M_{\rm pl}}X\,,\qquad
G_4 (\phi,X)=\frac{M_{\rm pl}^2}{2} e^{-2Q\phi/M_{\rm pl}}+
\frac{c_4}{M^6}e^{-\lambda_4 \phi/M_{\rm pl}}X^2\,,\qquad
G_5(\phi, X)=\frac{c_5}{M^9}e^{-\lambda_5 \phi/M_{\rm pl}}X^2\,,
\label{dilagali}
\ee
where $\lambda_{3,4,5}$ are dimensionless constants of 
the order of unity. 
The dilatonic coupling of the above form arises not only 
in low-energy effective string theory \cite{Cartier} but also in the 
conformal Galileon model characterized by a probe brane
moving in an Anti-de Sitter throat \cite{Rham}. 
In the following we study two different cases: 
(i) $c_5 \neq 0$, $c_3=0$, $c_4=0$, and 
(ii) $c_3 \neq 0$, $c_4 \neq 0$, $c_5=0$, separately.

\subsubsection{$c_5 \neq 0$, $c_3=0$, $c_4=0$}
\label{secnonzeroc5}

Since the terms such as 
$G_{5,\phi X}$ and $G_{5,\phi XX}$ in Eqs.~(\ref{mu5})
and (\ref{beta}) do not vanish for the function $G_5$ 
involving the field $\phi$, it seems to be possible for the term
$G_5(\phi, X)=c_5 e^{-\lambda_5 \phi/M_{\rm pl}}X^2/M^9$ 
alone to accommodate the Vainshtein mechanism.
In the following we study this possibility by assuming that 
$|\lambda_5|$ is of the order of 1.

{}From Eq.~(\ref{varadius}) the Vainshtein radius can be estimated 
as $r_V=(6|c_5 \lambda_5| Q^4 M_{\rm pl}^3r_g^4/M^9)^{1/10}$.
For $r$ smaller than $r_V$ we have $\xi_1/\xi_2 \simeq 2$ 
in Eq.~(\ref{fieldap}), so that the field derivative 
is the same as that derived in Sec.~\ref{concretemodel}, i.e., 
\be
\phi'_{\rm out}(r)=\frac{QM_{\rm pl}r_g}{r_V^2}\,.
\label{phi52}
\ee
The term $|4X(G_{5,X}+XG_{5,XX})/r^2|$ 
in Eq.~(\ref{mu4}) gets larger than the other term 
$|2G_{4,\phi}|$ for $r<r_V/\sqrt{3|\lambda_5 Q|} \approx r_V$. 
Thus in the regime $r\ll r_V$ we have that 
$\mu_4\simeq-1/(3\lambda_5 M_{\rm pl})$ and 
$\mu_5\simeq 2 Q M_{\rm pl} r_g/(r_V^2 r)$. 
The condition $|\mu_4|\rho_m(r)<|\mu_5|$ is
violated for
\be
\rho_m(r) r > 6 |Q\lambda_5| \frac{M_{\rm pl}^2 r_g}{r_V^2}
\approx \frac{M_{\rm pl}^2 r_g}{r_V^2}\,.
\ee
Around the surfaces ($r=r_s$) of the Sun and the Earth with 
$M^3 \sim M_{\rm pl}H_0^2$, 
the above inequality is satisfied for $\rho_m(r_s)$ larger than 
$10^{-17}$ g/cm$^3$ and $10^{-16}$ g/cm$^3$, respectively.

The solution (\ref{phi52}) is subject to change inside 
the star.
Around the very vicinity of the center of the star the term 
$\mu_4$ can be estimated as $\mu_4 \simeq -G_{4,\phi}r/(2G_4\beta)$
with $\beta \simeq r$, so that the field equation (\ref{phieq}) reads
$\square \phi \simeq Q\rho_m/M_{\rm pl}$.
Assuming that $\rho_m$ approaches a constant 
$\rho_c$ for $r \to 0$, the solution to this equation is given by 
$\phi'(r) \simeq Q\rho_c r/(3M_{\rm pl})$.
The last term in the square bracket of Eq.~(\ref{mu4}) dominates 
over the $2G_{4,\phi}$ term for 
$r>r_c \equiv \sqrt{81M^9M_{\rm pl}^5/(2 |c_5 Q^3| \rho_c^4)}$.
If $\rho_c \sim 100$ g/cm$^3$ and $M^3 \sim M_{\rm pl}H_0^2$, 
then $r_c \approx 10^{-35}$ cm. 
For the distance $r$ larger than $r_c$ the field equation 
(\ref{phieq}) is simplified as
\be
\frac{d}{dr}(r^2\phi')\simeq-\frac{\rho_m}{3\lambda_5 M_{\rm pl}}r^2 
+2r\phi'\,.
\ee
Under the approximation that $\rho_m$ is nearly constant 
inside the star, 
we obtain the following solution for $r_c \ll r \lesssim r_s$:
\be
\phi'_{\rm in}(r) \simeq-\frac{\rho_m}{3\lambda_5 M_{\rm pl}} r\,. 
\label{phiori5}
\ee
In order to match this with another solution (\ref{phi52}), we 
require the condition $\lambda_5 Q<0$.
Since the Schwarzschild radius can be estimated as 
$r_g \simeq \rho_m r_s^3/(3M_{\rm pl}^2)$ from Eq.~(\ref{rgdef}), 
the ratio between $\phi'_{\rm in}(r)$ and $\phi'_{\rm out}(r)$
around $r=r_s$ is approximately given by 
\be
\left| \frac{\phi'_{\rm in}(r_s)}{\phi'_{\rm out}(r_s)} \right|
\approx \frac{1}{|\lambda_5 Q|} \left( \frac{r_V}{r_s}
\right)^2 \approx  \left( \frac{r_V}{r_s}
\right)^2\,.
\ee
For the matching of two solutions we require that 
$r_V \approx r_s$, but the Vainshtein mechanism 
works outside the star only for $r_V \gg r_s$. 
Hence the interior solution (\ref{phiori5}) does not 
connect to the exterior solution (\ref{phi52}) 
that accommodates the Vainshtein mechanism. 
In other words, if we integrate the field equation outwards with 
the boundary condition $\phi'(0)=0$, 
the field derivative becomes too large to be compatible 
with local gravity constraints around the surface of the star. 
This is an example where the Vainshtein mechanism does 
not operate inside the star.

While the above discussion corresponds to the case of
nearly constant $\rho_m$ inside the body, we also solved
Eq.~(\ref{phieq}) numerically for the density profile (\ref{profile}).
We confirmed that the interior and exterior solutions given above 
do not match with each other for $r_V \gg r_s$.

\subsubsection{$c_3 \neq 0$, $c_4 \neq 0$, $c_5=0$}

This case corresponds to the extension of the covariant 
Galileon model studied in Sec.~\ref{c50}.
Using the solution (\ref{phila}) in the regime $r \gg r_V$, 
one can show that the term $|2(G_{3,\phi}+XG_{3,\phi X})|$ 
in Eq.~(\ref{varadius}) is of the order of $r_g/r_V$ and that the term 
$|4(-3G_{4,\phi X}-2XG_{4,\phi XX})\phi'(r_V)|$ is suppressed
by the factor $r_g/r_V$ relative to the last term of Eq.~(\ref{varadius}).
Then, the Vainshtein radius is practically the same 
as Eq.~(\ref{vacv}).

In the regime $r_g \ll r \ll r_V$ the field equation of 
motion (\ref{fieldap}) reads
\be
\phi''(r)+\frac{\phi'(r)}{2r} \left[ 1-\frac{3\alpha_{43}\phi'(r)}
{M^3 r}-5u (r) \right]^{-1} \left[ 1-u(r) \right] \simeq 0\,,
\label{phidd}
\ee
where $\alpha_{43}=c_4/c_3$ and 
$u(r)=\lambda_4 \alpha_{43}\phi'(r)^2/(M^3M_{\rm pl})$.
Integration of Eq.~(\ref{phidd}) gives 
\be
r\phi'(r)^2 \left[ 1-u(r) \right]^4-\frac{2\alpha_{43}}{M^3}
\phi'(r)^3 \left[ 1-\frac95 u(r)+\frac97 u^2(r)
-\frac13 u^3(r) \right]=C\,,
\label{imsol2}
\ee
where $C$ is an integration constant determined 
by substituting the solution $\phi'(r)=Q M_{\rm pl}r_g/r_V^2$ at $r=r_V$. 
In the limit $|\alpha_{43}| \ll 1$, the leading-order solution 
to Eq.~(\ref{imsol2}) is the same as (\ref{phiel}), i.e., 
$\phi'(r) \simeq QM_{\rm pl}r_g/(r_V^{3/2}r^{1/2})$.
In this case we have $u(r) \simeq (\lambda_4 Q \alpha_{43}/c_V)(r_g/r)$, 
so that the correction from the non-zero $\lambda_4$ to the leading-order 
solution is very small.
When $|\alpha_{43}| \sim 1$, the leading-order solution is given by 
$\phi'(r) \simeq QM_{\rm pl}r_g/r_V^2$ for most of $r$ smaller than $r_V$. 
Since $u(r) \simeq (\lambda_4 Q \alpha_{43}/c_V)(r_g/r_V)$ in this case, 
the correction is suppressed as well. 
In the limit $|\alpha_{43}| \gg 1$ we have $|u(r)| \gg 1$ and
hence Eq.~(\ref{imsol2}) reduces to 
$\phi'(r)^9 [r\phi'(r)+2M_{\rm pl}/(3\lambda_4)] \simeq 
2M_{\rm pl}\phi'(r_V)^9/(3\lambda_4)$. 
Here we used the relation $|r_V\phi'(r_V)| 
\ll |2M_{\rm pl}/(3\lambda_4)|$ to determine the 
integration constant. We then obtain the solution 
$\phi'(r) \simeq (QM_{\rm pl}r_g/r_V^2)
[1-\lambda_4 Qr_gr/(6r_V^2)]$, which shows 
that the correction from the non-zero $\lambda_4$ is very small.

Inside the star, the correction from the $\lambda_4$ term to 
the leading-order solution is also suppressed.
For the theory with $c_3=0$ and the density profile 
(\ref{profile}), the variable $y$ defined 
in Eq.~(\ref{xz}) obeys the following approximate equation 
\be
\frac{dy(x)}{dx} \simeq \frac{1}{4c_4} x 
\left[ Qx e^{\lambda_4 z(x)}+6c_4 b_1^3 e^{2Qz(x)} y(x) 
\right] e^{-x^2/\beta_t^2}\,,
\label{dyxeqs3}
\ee
which is valid for $b_2 \ll 1$.
Since the variable $z$ satisfies the same equation as
(\ref{dzx}), the variation of $z$ is very tiny for $b_2 \ll 1$
and hence $e^{\lambda_4 z(x)} \simeq 1$.
In this case, Eq.~(\ref{dyxeqs3}) reduces to Eq.~(\ref{dyxeqs}).
The similar property also holds for the theory with $c_4=0$.

Thus the model (\ref{dilagali}) with $c_5=0$ can successfully 
accommodate the Vainshtein mechanism.
We note that the Vainshtein mechanism is also at work 
for extended Galileons with dilatonic couplings 
characterized by the Lagrangians 
$G_3(\phi, X)=c_3 M^{1-4p_3} 
e^{-\lambda_3 \phi/M_{\rm pl}}X^{p_3}$ ($p_3 \geq 1$)
and $G_4 (\phi,X)=(M_{\rm pl}^2/2) e^{-2Q\phi/M_{\rm pl}}+
c_4M^{2-4p_4}e^{-\lambda_4 \phi/M_{\rm pl}}X^{p_4}$ ($p_4 \geq 2$).

\subsection{DBI Galileons with Gauss-Bonnet and other terms}

In higher-dimensional theories there appears a scalar degree of freedom 
associated with the size of compact space or with the position of a 
probe brane in large extra dimensions. 
In the set-up of a relativistic probe brane embedded in 
a five-dimensional bulk, de Rham and Tolley \cite{Rham} showed that all the 
Galileon self-interactions and its generalizations arise from the 
brane tension, induced curvature, 
and the Gibbons-Hawking-York boundary terms.
If we consider a Gauss-Bonnet term in a higher-dimensional 
space-time, the dimensional reduction on a compact space 
gives rise to a self-interaction $X \square \phi$ of a scalar field $\phi$ 
(corresponding to the size of the extra dimensions) as well as 
other interactions with $\phi$ \cite{Davis,Vainother}.
In order to accommodate such scenarios, let us consider the 
following four-dimensional action
\ba
S&=& \int d^4 x \sqrt{-g} \biggl[ \frac{M_{\rm pl}^2}{2}e^{-2Q\phi/M_{\rm pl}}R
-f_1(\phi)^{2} \mu^4 \biggl( \sqrt{1-\frac{2f_1(\phi)^{-1}X}{\mu^4}}-1 \biggr)
+f_2(\phi) \frac{X^2}{\mu^4} \nonumber \\
& &~~~~~~~~~~~~~~~~
+f_3(\phi) \frac{X}{M^3} \square \phi+f_4(\phi)c_{\rm GB}
R_{\rm GB}^2+f_5(\phi) \frac{1}{m^2} G^{\mu \nu}
\partial_{\mu} \phi \partial_{\nu} \phi \biggr]\,,
\label{actiontheory}
\ea
where $\mu$, $M$, $c_{\rm GB}$, $m$ are constants, and 
$R_{\rm GB}^2$ is the Gauss-Bonnet term defined by 
\be
R_{\rm GB}^2\equiv R^2-4\,R_{\mu\nu}R^{\mu\nu}
+R_{\alpha\beta\gamma\delta}R^{\alpha\beta\gamma\delta}\,.
\ee
For the functions $f_i(\phi)$ ($i=1,\cdots,5$) we assume that, 
without loss of generality, they 
are all equivalent to 
\be
f(\phi)=e^{-\lambda \phi/M_{\rm pl}}\,,
\ee
where $\lambda$ is a constant of the order of 1.
The second term in Eq.~(\ref{actiontheory})
corresponds to the DBI term appearing in the relativistic set-up 
of a probe brane moving in an Anti-de Sitter throat \cite{Rham}. 
The last four terms in Eq.~(\ref{actiontheory}) arise after the 
dimensional reduction of a higher-dimensional 
Gauss-Bonnet theory \cite{Vainother}
or as higher-order $\alpha'$-corrections to the low-energy 
bosonic string action \cite{Cartier}. In the case of 
$\alpha'$-corrections, the mass scales $\mu$, $M$, 
and $m$ are usually 
much higher than those related to dark energy.
In the following we do not put some restriction on 
the mass scales from the beginning, but we  
constrain those scales from the demand of 
realizing the Vainshtein mechanism.
In the action (\ref{actiontheory}) we can also take into 
account other dilatonic Galileon self-interactions, but those effects 
are similar to what we studied in Sec.~\ref{conformalsec}.
 
The Gauss-Bonnet coupling $f(\phi)R_{\rm GB}^2$ gives 
rise to the same equations of motion as those derived from 
the Horndeski's action (\ref{action}) for the choice
$K=8f^{(4)}(\phi) X^2 [3-\ln (X/\mu^4)]$, 
$G_3=4 f^{(3)}(\phi) X [7-3\ln (X/\mu^4)]$, 
$G_4=4 f^{(2)}(\phi) X [2-\ln (X/\mu^4)]$, and  
$G_5=-4 f^{(1)}(\phi) \ln (X/\mu^4)$,
where $f^{(n)}(\phi)\equiv d^n f/d\phi^n$ \cite{Koba11}. 
The last term in Eq.~(\ref{actiontheory}) is equivalent to 
$G_5(\phi)G^{\mu \nu} (\nabla_{\mu} \nabla_{\nu}\phi)$
with $G_{5}(\phi)=M_{\rm pl}f(\phi)/(\lambda m^2)$ after 
integration by parts\footnote{The model characterized by 
$G_5(\phi) \propto \phi$ corresponds to the one studied 
in Refs.~\cite{Einderi}.}.
In the language of the Horndeski's action (\ref{action}), 
the theory (\ref{actiontheory}) corresponds to 
\ba
K(\phi,X) &=& 
-f(\phi)^{2} \mu^4 \biggl( \sqrt{1-\frac{2f(\phi)^{-1} X}{\mu^4}}-1 \biggr)
+f(\phi)\frac{X^2}{\mu^4}
+8c_{\rm GB}f^{(4)}(\phi) X^2 \left[3-\ln \left(\frac{X}{\mu^4}\right)\right]\,,
\label{Kcom} \\
G_{3}(\phi,X) &=&
-f(\phi) \frac{X}{M^3}+4c_{\rm GB} f^{(3)}(\phi) X 
\left[7-3\ln \left(\frac{X}{\mu^4}\right)\right]\,,\\
G_{4}(\phi,X) &=&
\frac{M_{\rm pl}^2}{2}e^{-2Q\phi/M_{\rm pl}}
+4c_{\rm GB} f^{(2)}(\phi) X
\left[2-\ln \left(\frac{X}{\mu^4}\right)\right]\,,\\
G_{5}(\phi,X) &=& \frac{M_{\rm pl}}{\lambda m^2}f(\phi)
-4c_{\rm GB}  f^{(1)}(\phi) \ln \left( \frac{X}{\mu^4} 
\right)\,.
\ea

Provided that the conditions $|X/\mu^4| \ll 1$ and $|X/(M^3 M_{\rm pl})| \ll 1$ 
are satisfied, the function (\ref{Kcom})
reduces to the form 
$K(\phi,X) \simeq f(\phi)X+8c_{\rm GB}f^{(4)}(\phi) X^2 
[3-\ln (X/\mu^4 )]$.
The contribution of the Gauss-Bonnet term vanishes 
for the term $\beta$ in Eq.~(\ref{beta}). 
Then, the Vainshtein radius is known from 
Eq.~(\ref{varadius}) as
\be
r_V \simeq (4|Q|M_{\rm pl}r_g)^{1/3}/M\,,
\label{rVstring}
\ee
where we used Eq.~(\ref{phila}).
In the regime $r \gg r_V$ we recall that the solution $\phi'(r)=QM_{\rm pl}r_g/r^2$
is valid under several conditions presented in Sec.~\ref{rlarge}. 
Now we have $\beta \simeq f(\phi)r \simeq r$ 
for $r \gg r_V$, where the second approximate equality 
is valid for $r_V \gg r_g$ [i.e., equivalent to (\ref{con3la})].
For $r_V \gg r_g$ we have $|X/(M^3 M_{\rm pl})| \ll 1$. 
Provided that 
\be
|c_{\rm GB} X/M_{\rm pl}^4| \ll 1\,,\qquad 
|X/\mu^4| \ll 1\,,
\label{scon1}
\ee
both (\ref{con1la}) and (\ref{con2la}) are met in the regime $r \gg r_V$, 
so that $\mu_4 \simeq Q/M_{\rm pl}$ and 
$\mu_5 \simeq -12X(1-\phi'M^3 r/\mu^4)/(M^3 r^2)$. 
Under the condition
\be
\mu^4 \gg |Q|M^3 M_{\rm pl} \frac{r_g}{r_V}\,,
\label{scon2}
\ee
it follows that  $\mu_5 \simeq -12X/(M^3 r^2)$. 
Then, the distance $r_*$ at which $|\mu_4| \rho_m$ becomes 
the same order as $\mu_5$ can be estimated as 
$r_*/r_V \approx [M^3M_{\rm pl}/(|Q|\rho_m)]^{1/6}$.
For $M^3 \approx M_{\rm pl}H_0^2$ and $\rho_m$ close to $\rho_0$, 
$r_*$ is the same order as $r_V$.

In the regime $r_g \ll r \ll r_V$, the terms
in Eq.~(\ref{xi1}) are simply given by 
$\xi_1=-3re^{-\lambda \phi/M_{\rm pl}}/M^3$ and 
$\xi_2=-2r e^{-\lambda \phi/M_{\rm pl}}/M^3$.
Then the solution to the field equation (\ref{fieldap}), 
after matching at $r=r_V$, is 
\be
\phi'(r) \simeq \frac{QM_{\rm pl}r_g}{r_V^{3/2}r^{1/2}}\,.
\label{phidf}
\ee
Using this solution as well as the conditions (\ref{scon1}) and 
(\ref{scon2}), one can show that Eqs.~(\ref{con1sm}), (\ref{con2sm}), 
and (\ref{con3sm}) are satisfied. 
If the condition 
\be
m^2 \gg \frac{|\phi''(r)|}{|Q|M_{\rm pl}}\,,
\label{scon3}
\ee
is met in addition to (\ref{scon1}) and 
(\ref{scon2}), Eqs.~(\ref{eq:00}) and (\ref{eq:11}) of the 
gravitational potentials approximately reduce to the same equations 
as (\ref{G3Phi}) and (\ref{G3Psi}) respectively.
Then the gravitational potentials in the regime 
$r_g \ll r \ll r_V$ are given by Eq.~(\ref{Psiso2}), 
so that the fifth force is suppressed
deep inside the Vainshtein radius.

Inside the star ($r<r_s$) the solution to the field 
equation is subject to change. 
As long as the conditions (\ref{scon1}), (\ref{scon2}), and 
(\ref{scon3}) are satisfied, the situation is similar to 
what we studied in Sec.~\ref{exgali} for $p_3=1$.
Around the radius of the star the solution (\ref{phidf}) 
smoothly connects to another solution 
$\phi'(r) \propto r$ (see Fig.~\ref{fig3}).

Since $|\phi'(r)|$ reaches a maximum around the surface of the star, 
we can substitute the values $\phi'(r_s)$ and $\phi''(r_s)$ into 
Eqs.~(\ref{scon1}) and (\ref{scon3}) to derive the bounds of $c_{\rm GB}$, 
$\mu$, and $m$, as 
\be
|c_{\rm GB}| \ll \frac{r_V^3r_s M_{\rm pl}^2}{Q^2 r_g^2}\,,\qquad
\mu \gg \left( \frac{Q^2 M_{\rm pl}^2 r_g^2}{r_V^3 r_s} \right)^{1/4}\,,
\qquad
m \gg \left( \frac{r_g^2}{r_V^3r_s^3} \right)^{1/4}\,.
\label{condi}
\ee
The condition (\ref{scon2}) gives a weaker bound on $\mu$
than the second of Eq.~(\ref{condi}).
If we demand that the experimental bound of the post-Newtonian parameter 
(i.e., $|\gamma-1| \approx  Q^2(r/r_V)^{3/2}<2.3 \times 10^{-5}$) is 
satisfied up to the scales 
$r=10\,$Au\,$ \approx 10^{14}$\,cm, then $r_V$ needs to be 
larger than $10^{17}$\,cm for $|Q|={\cal O}(1)$.
On using Eq.~(\ref{rVstring}), this corresponds to the mass 
scale $M \lesssim 10^{-18}$~GeV.
In the case of the Sun with the Vainshtein radius $r_V=10^{20}$\,cm, 
for example, the conditions (\ref{condi}) translate to 
$|c_{\rm GB}| \ll 10^{124}$, $\mu \gg 10^{-13}$\,GeV, and 
$m \gg 10^{-34}$\,GeV.
In particular, the Gauss-Bonnet coupling with $|c_{\rm GB}| \sim 1$ 
does not give rise to any modification to the Vainshtein mechanism.
It is worthy of mentioning that in the field equation (\ref{phieq}) the effect of
the Gauss-Bonnet coupling appears only in the $G_4$ term 
of Eq.~(\ref{mu4}).

\section{Conclusions}
\label{consec} 

In this paper we have studied the Vainshtein mechanism in the most general 
second-order scalar-tensor theories given by the action (\ref{action}). 
We derived the full equations of motion (\ref{eq:00})-(\ref{continuity})
for a spherically symmetric metric (\ref{line}) characterized 
by two gravitational potentials $\Psi$ and $\Phi$.
Under the weak gravity approximation the equations of 
motion for the field $\phi$ and the gravitational potential $\Psi$ 
reduce to fairly simple forms (\ref{phieq}) and (\ref{Poimo}), respectively.
These equations can be used to study the Vainshtein screening 
effect as well as the chameleon and symmetron mechanisms.

In the presence of a non-minimal coupling $e^{-2Q\phi/M_{\rm pl}}$ 
with the Ricci scalar $R$, we clarify conditions under which  
the Vainshtein mechanism operates due to the field non-linear 
self-interactions. The Vainshtein radius $r_V$ 
is implicitly given by the formula (\ref{varadius}), from which 
$r_V$ is known explicitly for a given model.
For the distance $r$ larger than $r_V$ the non-linear field 
self-interactions are suppressed, so that the solution to 
Eq.~(\ref{phieq}) is $\phi'(r)=QM_{\rm pl}r_g/r^2$. For 
the validity of this solution we require that all the conditions
(\ref{con1la}), (\ref{con2la}), (\ref{con4la}), and (\ref{con5la}) 
are satisfied. For the distance characterized by $r_g \ll r \ll r_V$, 
the field equation (\ref{phieq}) reduces to (\ref{fieldap}) under 
the conditions (\ref{con1sm}), (\ref{con2sm}), and (\ref{con3sm}).
This is the regime in which the Vainshtein mechanism 
works to suppress the propagation of the fifth force. 
Inside a spherically symmetric body ($r<r_s$),  
the solution is different from that 
in the regime $r_g \ll r \ll r_V$.
For the smooth matching of two solutions the Vainshtein 
mechanism needs to be at work inside the body as well.

The covariant Galileon model characterized by 
$G_4=M_{\rm pl}^2 e^{-2Q\phi/M_{\rm pl}}/2+c_4X^2/M^6$ 
and $G_3=G_5=0$ is a prototype that accommodates
the Vainshtein mechanism successfully.
In this model there is the solution 
$\phi'(r)=QM_{\rm pl}r_g/r_V^2=$~constant for the 
distance $r_g \ll r \ll r_V$. 
In this regime the gravitational potentials are given by 
Eq.~(\ref{Psiso1}), in which case local gravity constraints 
are well satisfied.
In Sec.~\ref{concretemodel} we confirmed that all the conditions 
to derive the solutions (\ref{phila}) and (\ref{phiin2}) are 
consistently satisfied.
For the star with a constant density there is 
the solution $\phi'(r) \propto r$ with which the Vainshtein 
mechanism is at work inside the body.
For the varying density characterized by the profile 
(\ref{profile}) we numerically showed that the interior
solution smoothly connects with the exterior
solution (\ref{phiin2}).
This result is insensitive to the choice of the density 
profile, so that the Vainshtein mechanism 
operates successfully both outside and inside the body.

In Sec.~\ref{galisec} we studied the covariant Galileon model
in which all the non-linear derivative terms in $G_{3,4,5}$ exist.
In the absence of the term $G_5=c_5 X^2/M^9$ we showed 
that the Vainshtein mechanism is at work to suppress 
the fifth force inside the Vainshtein radius.
However, if the term $G_5=c_5 X^2/M^9$ is present, 
this modifies the solution of the field equation (\ref{phieq}) 
due to the appearance of the term $4X(G_{5,X}+XG_{5,XX})/r^2$
in Eq.~(\ref{mu4}). For the model with $|c_4| \sim 1$ and $|c_3| \ll 1$, 
unless the coefficient $c_5$ satisfies the condition (\ref{c5condi}), 
the Vainshtein mechanism does not operate inside the star 
and hence there is a problem of matching the solutions 
around the surface of the body.
For the model with $|c_3| \sim 1$ and $|c_4| \ll 1$, 
unless $|c_5|$ is much smaller than 1, we showed that 
local gravity constraints are not satisfied within the solar system.
These results are consistent with those of 
Kimura {\it et al.} \cite{Kimura} and Koyama {\it et al.} \cite{KNT}, but 
we derived the conditions for the success of the Vainshtein mechanism
more precisely in the presence of all the covariant Galileon terms.

In Sec.~\ref{appsec} we applied our results to several models 
such as extended Galileons, covariant Galileons with dilatonic 
couplings, and DBI Galileons with Gauss-Bonnet and 
other terms. As long as the non-linear derivative terms coupled to 
the Einstein tensor 
($G_5=c_5M^{-1-4p_5}e^{-\lambda_5 \phi/M_{\rm pl}}X^{p_5}$ with $p_5 \geq 2$) 
do not dominate over other non-linear field self-interactions, we showed 
that the Vainshtein mechanism is at work both inside and outside 
the star. In short, the dominance of the terms such as 
$G_3=c_3 M^{1-4p_3} 
e^{-\lambda_3 \phi/M_{\rm pl}}X^{p_3}$ ($p_3 \geq 1$)
and $G_4=(M_{\rm pl}^2/2) e^{-2Q\phi/M_{\rm pl}}+
c_4M^{2-4p_4}e^{-\lambda_4 \phi/M_{\rm pl}}X^{p_4}$ ($p_4 \geq 2$)
signals the success of the Vainshtein mechanism. 
The contributions of the Gauss-Bonnet term, the higher-order terms 
of $X$, and the non-minimal coupling to the Einstein tensor are 
suppressed under the condition (\ref{condi}), in which case 
the success of the Vainshtein screening is not modified.

Our analysis in this paper was carried out on the spherically symmetric 
background, so it is not valid on the cosmologically large scales at which 
the time variations of physical quantities are non-negligible. 
In Ref.~\cite{timeva} it was shown that the time variation of 
the Newton ``constant'' $G_N$ can put tight constraints on 
scalar-tensor theories when the matter-scalar coupling 
is of the order of unity. 
In order to address this point, we need to discuss solutions of 
the field equations in the spherically symmetric configurations 
on the time-dependent 
cosmological background (along the line of Ref.~\cite{Kimura}).
It will be certainly of interest to study whether there exist 
dark energy models based on the Horndeski's theory which can 
be compatible with both local gravity constraints and the bounds 
of the time variation of $G_N$. 
We leave this for a future work.

\section*{ACKNOWLEDGEMENTS}
\label{acknow} 

R.\ K.\ and S.\ T.\ are supported by the Scientific Research Fund of the JSPS 
(Nos. 24 $\cdot$ 6770 and 24540286). S.\ T.\ also thanks financial support from 
Scientific Research on Innovative Areas (No.~21111006).



\begin{thebibliography}{10}

\bibitem{review} 
T.~P.~Sotiriou and V.~Faraoni,
Rev.\ Mod.\ Phys.\  {\bf 82}, 451 (2010)
[arXiv:0805.1726 [gr-qc]];\\
A.~De Felice and S.~Tsujikawa,
Living Rev.\ Rel.\  {\bf 13}, 3 (2010)
[arXiv:1002.4928 [gr-qc]];\\
S.~Tsujikawa,
Lect.\ Notes Phys.\  {\bf 800}, 99 (2010)
[arXiv:1101.0191 [gr-qc]];\\
T.~Clifton, P.~G.~Ferreira, A.~Padilla and C.~Skordis,
Phys.\ Rept.\  {\bf 513}, 1 (2012)
[arXiv:1106.2476 [astro-ph.CO]].

\bibitem{Star80} 
A.~A.~Starobinsky,
Phys.\ Lett.\ B {\bf 91}, 99 (1980).

\bibitem{chame} 
J.~Khoury and A.~Weltman,
Phys.\ Rev.\ Lett.\  {\bf 93}, 171104 (2004)
[astro-ph/0309300];
Phys.\ Rev.\ D {\bf 69}, 044026 (2004)
[astro-ph/0309411].

\bibitem{fRlocal} 
I.~Navarro and K.~Van Acoleyen,
JCAP {\bf 0702}, 022 (2007)
[gr-qc/0611127];\\ 
T.~Faulkner, M.~Tegmark, E.~F.~Bunn and Y.~Mao,
Phys.\ Rev.\ D {\bf 76}, 063505 (2007)
[astro-ph/0612569];\\
W.~Hu and I.~Sawicki,
Phys.\ Rev.\ D {\bf 76}, 064004 (2007)
[arXiv:0705.1158 [astro-ph]];\\
S.~Capozziello and S.~Tsujikawa,
Phys.\ Rev.\ D {\bf 77}, 107501 (2008)
[arXiv:0712.2268 [gr-qc]].

\bibitem{Yoko} 
S.~Tsujikawa, K.~Uddin, S.~Mizuno, R.~Tavakol 
and J.~'i.~Yokoyama,
Phys.\ Rev.\ D {\bf 77}, 103009 (2008)
[arXiv:0803.1106 [astro-ph]];\\
R.~Gannouji, B.~Moraes, D.~F.~Mota, D.~Polarski, S.~Tsujikawa and H.~A.~Winther,
Phys.\ Rev.\ D {\bf 82}, 124006 (2010)
[arXiv:1010.3769 [astro-ph.CO]].

\bibitem{symmetron} 
K.~Hinterbichler and J.~Khoury,
Phys.\ Rev.\ Lett.\  {\bf 104}, 231301 (2010)
[arXiv:1001.4525 [hep-th]];\\
K.~Hinterbichler, J.~Khoury, A.~Levy and A.~Matas,
Phys.\ Rev.\ D {\bf 84}, 103521 (2011)
[arXiv:1107.2112 [astro-ph.CO]].

\bibitem{Vainshtein} 
A.~I.~Vainshtein, 
Phys.\ Lett.\ B \textbf{39}, 393 (1972).

\bibitem{Pauli} 
M.~Fierz, Helv.\ Phys.\ Acta \textbf{12}, 3 (1939);\\
M.~Fierz and W.~Pauli, Proc.\ Roy.\ Soc. \ Lond. A 
\textbf{173}, 211 (1939).

\bibitem{DVZ} 
H.~van Dam and M.\ J.\ G.~Veltman, Nucl.\ Phys.\, B \textbf{22}, 397 (1970); \\
V.\ I.~Zakharov, JETP Lett. \ \textbf{12}, 312 (1970); \\
Y.~Iwasaki, 
Phys.\ Rev.\ \textbf{D2}, 2255-2256 (1970).

\bibitem{DGP} 
G.~R.~Dvali, G.~Gabadadze and M.~Porrati,
Phys.\ Lett.\ B {\bf 485}, 208 (2000)
[hep-th/0005016].

\bibitem{DGPnon} 
C.~Deffayet, G.~R.~Dvali and G.~Gabadadze,
Phys.\ Rev.\ D {\bf 65}, 044023 (2002)
[astro-ph/0105068].

\bibitem{DDGV} 
C.~Deffayet, G.~R.~Dvali, G.~Gabadadze and A.~I.~Vainshtein,
Phys.\ Rev.\ D {\bf 65}, 044026 (2002)
[hep-th/0106001].

\bibitem{Nico} 
A.~Nicolis and R.~Rattazzi,
JHEP {\bf 0406}, 059 (2004)
[hep-th/0404159].

\bibitem{Silva} 
K.~Koyama and F.~P.~Silva,
Phys.\ Rev.\ D {\bf 75}, 084040 (2007)
[hep-th/0702169].

\bibitem{Nicolis} 
A.~Nicolis, R.~Rattazzi and E.~Trincherini,
Phys.\ Rev.\ \textbf{D79}, 064036 (2009).

\bibitem{Deffayet} 
C.~Deffayet, G.~Esposito-Farese and A.~Vikman,
Phys.\ Rev.\ D {\bf 79}, 084003 (2009)
[arXiv:0901.1314 [hep-th]];\\
C.~Deffayet, S.~Deser and G.~Esposito-Farese,
Phys.\ Rev.\ D {\bf 80}, 064015 (2009)
[arXiv:0906.1967 [gr-qc]].

\bibitem{GALcosmo} 
R.~Gannouji and M.~Sami,
Phys.\ Rev.\ D {\bf 82}, 024011 (2010);\\
C.~Deffayet, O.~Pujolas, I.~Sawicki and A.~Vikman,
JCAP {\bf 1010}, 026 (2010)
[arXiv:1008.0048 [hep-th]];\\
A.~Ali, R.~Gannouji and M.~Sami, 
Phys.\ Rev.\ \textbf{D82}, 103015 (2010);\\ 
S.~Nesseris, A.~De Felice and S.~Tsujikawa, 
Phys.\ Rev.\ \textbf{D82}, 124054 (2010);\\ 
D.~F.~Mota, M.~Sandstad and T.~Zlosnik, 
JHEP \textbf{1012}, 051 (2010); \\
A.~De Felice, R.~Kase and S.~Tsujikawa,
Phys.\ Rev.\ \textbf{D83}, 043515 (2011);\\ 
C.~de Rham and L.~Heisenberg,
Phys.\ Rev.\ \textbf{D84}, 043503 (2011);\\
S.~Appleby and E.~V.~Linder,
JCAP {\bf 1203}, 043 (2012)
[arXiv:1112.1981 [astro-ph.CO]];
JCAP {\bf 1208}, 026 (2012)
[arXiv:1204.4314 [astro-ph.CO]];\\
A.~Barreira, B.~Li, C.~M.~Baugh and S.~Pascoli,
Phys.\ Rev.\ D {\bf 86}, 124016 (2012)
[arXiv:1208.0600 [astro-ph.CO]];\\
H.~Okada, T.~Totani and S.~Tsujikawa,
Phys.\ Rev.\ D {\bf 87}, 103002 (2013)
[arXiv:1208.4681 [astro-ph.CO]];\\
N.~Bartolo, E.~Bellini, D.~Bertacca and S.~Matarrese,
JCAP {\bf 1303}, 034 (2013)
[arXiv:1301.4831 [astro-ph.CO]];\\
P.~Creminelli {\it et.al.},
JHEP {\bf 1302}, 006 (2013)
[arXiv:1209.3768 [hep-th]];\\
J.~Neveu {\it et. al.},
arXiv:1302.2786 [gr-qc];\\
A.~Barreira, B.~Li, A.~Sanchez, C.~M.~Baugh and S.~Pascoli,
Phys.\ Rev.\ D {\bf 87}, 103511 (2013)
[arXiv:1302.6241 [astro-ph.CO]];
arXiv:1306.3219 [astro-ph.CO].

\bibitem{DTPRL} 
A.~De Felice and S.~Tsujikawa,
Phys.\ Rev.\ Lett.\  {\bf 105}, 111301 (2010);
Phys.\ Rev.\ D {\bf 84}, 124029 (2011).

\bibitem{Rham} 
C.~de Rham and A.~J.~Tolley,
JCAP {\bf 1005}, 015 (2010)
[arXiv:1003.5917 [hep-th]].

\bibitem{Trodden}
K.~Hinterbichler, M.~Trodden and D.~Wesley,
Phys.\ Rev.\ D {\bf 82}, 124018 (2010)
[arXiv:1008.1305 [hep-th]];\\
G.~Goon, K.~Hinterbichler and M.~Trodden,
JCAP {\bf 1107}, 017 (2011)
[arXiv:1103.5745 [hep-th]]; 
Phys.\ Rev.\ Lett.\  {\bf 106}, 231102 (2011)
[arXiv:1103.6029 [hep-th]];\\
J.~Khoury, J.~-L.~Lehners and B.~A.~Ovrut,
Phys.\ Rev.\ D {\bf 84}, 043521 (2011)
[arXiv:1103.0003 [hep-th]];\\
M.~Trodden and K.~Hinterbichler,
Class.\ Quant.\ Grav.\  {\bf 28}, 204003 (2011)
[arXiv:1104.2088 [hep-th]];\\
G.~Goon, K.~Hinterbichler, A.~Joyce and M.~Trodden,
JHEP {\bf 1206}, 004 (2012)
[arXiv:1203.3191 [hep-th]];\\
A.~Padilla and V.~Sivanesan,
JHEP {\bf 1304}, 032 (2013)
[arXiv:1210.4026 [gr-qc]];\\
P.~de Fromont, C.~de Rham, L.~Heisenberg and A.~Matas,
arXiv:1303.0274 [hep-th];\\
F.~Farakos, C.~Germani and A.~Kehagias,
arXiv:1306.2961 [hep-th].

\bibitem{Ostro} 
M.~Ostrogradski, 
Mem.\ Ac.\ St.\ Petersbourg VI 4, 385 (1850).

\bibitem{Horndeski} 
G.~W.~Horndeski, 
Int.\ J.\ Theor.\ Phys.\ 10, 363-384 (1974).

\bibitem{Deffayet11} 
C.~Deffayet, X.~Gao, D.~A.~Steer and G.~Zahariade,
Phys.\ Rev.\ D {\bf 84}, 064039 (2011)
[arXiv:1103.3260 [hep-th]].

\bibitem{Charmo} 
C.~Charmousis, E.~J.~Copeland, A.~Padilla and P.~M.~Saffin,
Phys.\ Rev.\ Lett.\  {\bf 108}, 051101 (2012)
[arXiv:1106.2000 [hep-th]].

\bibitem{Koba11} 
T.~Kobayashi, M.~Yamaguchi and J.~Yokoyama,
Prog.\ Theor.\ Phys.\  {\bf 126}, 511 (2011)
[arXiv:1105.5723 [hep-th]].

\bibitem{Hoinf} 
S.~Weinberg,
Phys.\ Rev.\ D {\bf 77}, 123541 (2008)
[arXiv:0804.4291 [hep-th]];\\
C.~Cheung, P.~Creminelli, A.~L.~Fitzpatrick, J.~Kaplan and L.~Senatore,
JHEP {\bf 0803}, 014 (2008)
[arXiv:0709.0293 [hep-th]].

\bibitem{Hoda} 
G.~Gubitosi, F.~Piazza and F.~Vernizzi,
JCAP {\bf 1302}, 032 (2013)
[arXiv:1210.0201 [hep-th]].

\bibitem{Fedo} 
J.~Gleyzes, D.~Langlois, F.~Piazza and F.~Vernizzi,
arXiv:1304.4840 [hep-th].

\bibitem{Horncosmo} 
E.~J.~Copeland, A.~Padilla and P.~M.~Saffin,
JCAP {\bf 1212}, 026 (2012)
[arXiv:1208.3373 [hep-th]];\\
A.~De Felice, T.~Kobayashi and S.~Tsujikawa,
Phys.\ Lett.\ B {\bf 706}, 123 (2011)
[arXiv:1108.4242 [gr-qc]];\\
L.~Amendola, M.~Kunz, M.~Motta, I.~D.~Saltas and I.~Sawicki,
Phys.\ Rev.\ D {\bf 87}, 023501 (2013)
[arXiv:1210.0439 [astro-ph.CO]].

\bibitem{exGAL} 
A.~De Felice and S.~Tsujikawa,
JCAP {\bf 1202}, 007 (2012)
[arXiv:1110.3878 [gr-qc]]; JCAP {\bf 1203}, 025 (2012) [arXiv:1112.1774 [astro-ph.CO]].

\bibitem{KY} 
R.~Kimura and K.~Yamamoto,
JCAP {\bf 1104}, 025 (2011)
[arXiv:1011.2006 [astro-ph.CO]].

\bibitem{VainARS}
A.~De Felice, R.~Kase and S.~Tsujikawa,
Phys.\ Rev.\ D {\bf 85}, 044059 (2012)
[arXiv:1111.5090 [gr-qc]].

\bibitem{Kimura} 
R.~Kimura, T.~Kobayashi and K.~Yamamoto,
Phys.\ Rev.\ D {\bf 85}, 024023 (2012)
[arXiv:1111.6749 [astro-ph.CO]].

\bibitem{Chow} 
N.~Chow and J.~Khoury,
Phys.\ Rev.\ D {\bf 80}, 024037 (2009)
[arXiv:0905.1325 [hep-th]].

\bibitem{Burrage} 
C.~Burrage and D.~Seery,
JCAP {\bf 1008}, 011 (2010)
[arXiv:1005.1927 [astro-ph.CO]].

\bibitem{Babichev} 
E.~Babichev, C.~Deffayet and R.~Ziour,
JHEP {\bf 0905}, 098 (2009)
[arXiv:0901.0393 [hep-th]];
Phys.\ Rev.\ Lett.\  {\bf 103}, 201102 (2009)
[arXiv:0907.4103 [gr-qc]];
Phys.\ Rev.\ D {\bf 82}, 104008 (2010)
[arXiv:1007.4506 [gr-qc]];\\
E.~Babichev and C.~Deffayet,
arXiv:1304.7240 [gr-qc].

\bibitem{BBD} 
P.~Brax, C.~Burrage and A.~-C.~Davis,
JCAP {\bf 1109}, 020 (2011)
[arXiv:1106.1573 [hep-ph]];
JCAP {\bf 1301}, 020 (2013)

\bibitem{Tana} 
N.~Kaloper, A.~Padilla and N.~Tanahashi,
JHEP {\bf 1110}, 148 (2011)
[arXiv:1106.4827 [hep-th]].

\bibitem{Wesley} 
C.~de Rham, A.~J.~Tolley and D.~H.~Wesley,
Phys.\ Rev.\ D {\bf 87}, 044025 (2013)
[arXiv:1208.0580 [gr-qc]].

\bibitem{Hiramatsu} 
T.~Hiramatsu, W.~Hu, K.~Koyama and F.~Schmidt,
Phys.\ Rev.\ D {\bf 87}, 063525 (2013)
[arXiv:1209.3364 [hep-th]];\\
A.~V.~Belikov and W.~Hu,
arXiv:1212.0831 [gr-qc];\\
B.~Li, G.~-B.~Zhao and K.~Koyama,
JCAP {\bf 1305}, 023 (2013)
[arXiv:1303.0008 [astro-ph.CO]].

\bibitem{Chu} 
Y.~-Z.~Chu and M.~Trodden,
Phys.\ Rev.\ D {\bf 87}, 024011 (2013)
[arXiv:1210.6651 [astro-ph.CO]];\\
M.~Andrews, Y.~-Z.~Chu and M.~Trodden,
arXiv:1305.2194 [astro-ph.CO].

\bibitem{Narikawa} 
T.~Narikawa, T.~Kobayashi, D.~Yamauchi and R.~Saito,
Phys.\  Rev.\ D {\bf 87}, 124006 (2013)
[arXiv:1302.2311 [astro-ph.CO]].

\bibitem{KNT} 
K.~Koyama, G.~Niz and G.~Tasinato,
arXiv:1305.0279 [hep-th].

\bibitem{Bere} 
L.~Berezhiani, G.~Chkareuli and G.~Gabadadze,
arXiv:1302.0549 [hep-th];
L.~Berezhiani, G.~Chkareuli, C.~de Rham, 
G.~Gabadadze and A.~J.~Tolley,
arXiv:1305.0271 [hep-th]. 

\bibitem{stpapers}
A.~V.~Frolov,
Phys.\ Rev.\ Lett.\  {\bf 101}, 061103 (2008)
[arXiv:0803.2500 [astro-ph]];\\
T.~Kobayashi and K.~-i.~Maeda,
Phys.\ Rev.\ D {\bf 78}, 064019 (2008)
[arXiv:0807.2503 [astro-ph]];\\
S.~Tsujikawa, T.~Tamaki and R.~Tavakol,
JCAP {\bf 0905}, 020 (2009)
[arXiv:0901.3226 [gr-qc]];\\
A.~Upadhye and W.~Hu,
Phys.\ Rev.\ D {\bf 80}, 064002 (2009)
[arXiv:0905.4055 [astro-ph.CO]];\\
E.~Babichev and D.~Langlois,
Phys.\ Rev.\ D {\bf 81}, 124051 (2010)
[arXiv:0911.1297 [gr-qc]].

\bibitem{Gasperini} 
M.~Gasperini and G.~Veneziano,
Astropart.\ Phys.\  {\bf 1}, 317 (1993)
[hep-th/9211021];\\
M.~Gasperini and G.~Veneziano,
Phys.\ Rept.\  {\bf 373}, 1 (2003)
[hep-th/0207130].

\bibitem{runaway} 
M.~Gasperini, F.~Piazza and G.~Veneziano,
Phys.\ Rev.\ D {\bf 65}, 023508 (2002)
[gr-qc/0108016];\\
T.~Damour, F.~Piazza and G.~Veneziano,
Phys.\ Rev.\ Lett.\  {\bf 89}, 081601 (2002)
[gr-qc/0204094];\\
F.~Piazza and S.~Tsujikawa,
JCAP {\bf 0407}, 004 (2004)
[hep-th/0405054].

\bibitem{Brans} 
C.~Brans and R.~H.~Dicke, 
Phys.\ Rev.\ \textbf{124}, 925 (1961).

\bibitem{Ohanlon} 
J.~O'Hanlon, Phys.\ Rev.\ Lett.\ \textbf{29},
137 (1972); \\
T.~Chiba, 
Phys.\ Lett.\ \textbf{B575}, 1-3 (2003).

\bibitem{Maeda} 
K.~-i.~Maeda,
Phys.\ Rev.\ D {\bf 39}, 3159 (1989).

\bibitem{DMota} 
P.~Brax, C.~van de Bruck, A.~-C.~Davis, J.~Khoury and A.~Weltman,
Phys.\ Rev.\ D {\bf 70}, 123518 (2004)
[astro-ph/0408415];\\
D.~F.~Mota and D.~J.~Shaw,
Phys.\ Rev.\ Lett.\  {\bf 97}, 151102 (2006)
[hep-ph/0606204];\\
P.~Brax, C.~van de Bruck, A.~-C.~Davis, D.~F.~Mota and D.~J.~Shaw,
Phys.\ Rev.\ D {\bf 76}, 124034 (2007)
[arXiv:0709.2075 [hep-ph]];\\
T.~Tamaki and S.~Tsujikawa,
Phys.\ Rev.\ D {\bf 78}, 084028 (2008)
[arXiv:0808.2284 [gr-qc]].

\bibitem{symap} 
P.~Brax, C.~van de Bruck, A.~-C.~Davis, B.~Li, B.~Schmauch and D.~J.~Shaw,
Phys.\ Rev.\ D {\bf 84}, 123524 (2011)
[arXiv:1108.3082 [astro-ph.CO]]; \\
A.~-C.~Davis, B.~Li, D.~F.~Mota and H.~A.~Winther,
Astrophys.\ J.\  {\bf 748}, 61 (2012)
[arXiv:1108.3081 [astro-ph.CO]];\\
P.~Brax and A.~-C.~Davis,
Phys.\ Lett.\ B {\bf 707}, 1 (2012)
[arXiv:1109.0468 [hep-ph]];\\
A.~Upadhye,
Phys.\ Rev.\ Lett.\  {\bf 110}, 031301 (2013)
[arXiv:1210.7804 [hep-ph]].

\bibitem{Davis} 
L.~Amendola, C.~Charmousis and S.~C.~Davis,
JCAP {\bf 0612}, 020 (2006)
[hep-th/0506137].

\bibitem{Vainother}
R.~Gannouji and M.~Sami,
Phys.\ Rev.\ D {\bf 85}, 024019 (2012)
[arXiv:1107.1892 [gr-qc]].

\bibitem{Mukhanov} 
C.~Armendariz-Picon, T.~Damour and V.~F.~Mukhanov,
Phys.\ Lett.\ B {\bf 458}, 209 (1999)
[hep-th/9904075]; \\
T.~Chiba, T.~Okabe and M.~Yamaguchi,
Phys.\ Rev.\ D {\bf 62}, 023511 (2000)
[astro-ph/9912463];\\
C.~Armendariz-Picon, V.~F.~Mukhanov and P.~J.~Steinhardt,
Phys.\ Rev.\ Lett.\  {\bf 85}, 4438 (2000)
[astro-ph/0004134].
 
\bibitem{Will} 
C.~M.~Will, 
Living Rev.\ Rel.\ \textbf{9}, 3 (2005). 
  
\bibitem{Cartier} 
R.~R.~Metsaev and A.~A.~Tseytlin,
Nucl.\ Phys.\ B {\bf 293}, 385 (1987);\\
M.~Gasperini, M.~Maggiore and G.~Veneziano,
Nucl.\ Phys.\ B {\bf 494}, 315 (1997)
[hep-th/9611039];\\
C.~Cartier, J.~-c.~Hwang and E.~J.~Copeland,
Phys.\ Rev.\ D {\bf 64}, 103504 (2001)
[astro-ph/0106197].

\bibitem{Einderi}
C.~Germani and A.~Kehagias,
Phys.\ Rev.\ Lett.\  {\bf 105}, 011302 (2010)
[arXiv:1003.2635 [hep-ph]];
Phys.\ Rev.\ Lett.\  {\bf 106}, 161302 (2011)
[arXiv:1012.0853 [hep-ph]];\\
S.~Tsujikawa,
Phys.\ Rev.\ D {\bf 85}, 083518 (2012)
[arXiv:1201.5926 [astro-ph.CO]].
 
\bibitem{timeva} 
E.~Babichev, C.~Deffayet and G.~Esposito-Farese,
Phys.\ Rev.\ Lett.\  {\bf 107}, 251102 (2011)
[arXiv:1107.1569 [gr-qc]]. 
 
\end{thebibliography}
\end{document}